\documentclass[a4paper,11pt]{article}
\usepackage[utf8]{inputenc}
\usepackage{jheppub,float}
\usepackage[normalem]{ulem}
\usepackage{xcolor}
\usepackage{amsmath,bm}
\usepackage{framed}
\usepackage{varioref}
\colorlet{shadecolor}{lightgray}
\usepackage{hyperref}
\usepackage{mathrsfs}
\usepackage{parskip}
\usepackage{xcolor}

\labelformat{equation}{Eq.~(#1)} 
\def \be {\begin{equation}}
\def \ee {\end{equation}}
\def \bea {\begin{eqnarray}}
\def \eea {\end{eqnarray}}
\def \nn {\nonumber}
\DeclareMathOperator{\arctanh}{arctanh}

\author[a,b]{Karan Fernandes,}
\author[a,b]{Feng-Li Lin}

\affiliation[a]{Department of Physics,
National Taiwan Normal University, Taipei, 11677, Taiwan}
\affiliation[b]{Center of Astronomy and Gravitation, National Taiwan Normal University, Taipei 11677, Taiwan}
\emailAdd{karanfernandes86@gmail.com, fengli.lin@gmail.com}
\title{Next-to-eikonal corrected double graviton dressing and gravitational wave observables at ${\cal O}(G^2)$}
\abstract{Following a recent proposal to describe inelastic eikonal scattering processes in terms of gravitationally dressed elastic eikonal amplitudes, we motivate a collinear double graviton dressing and investigate its properties. This is derived from a generalized Wilson line operator in the worldline formalism by integrating over fluctuations of the eikonal trajectories of external particles in gravitationally interacting theories. The dressing can be expressed as a product of exponential terms --  a coherent piece with contributions to all odd orders in the gravitational coupling constant and a term quadratic in graviton modes, with the former providing classical gravitational wave observables. In particular, the coherent dressing involves $\mathcal{O}(\kappa^3)$ subleading double graviton corrections to the Weinberg soft factor. We use this dressing to derive expressions for the waveform, radiative momentum spectrum and angular momentum. In a limiting case of the waveform, we derive the nonlinear memory effect resulting from the emission of nearly soft gravitons from a scattering process.}
\begin{document}
\maketitle
\flushbottom
\section{Introduction}

Scattering amplitude and effective field theory techniques have led to several advances in post-Minkowski (PM) results for binary gravitationally interacting systems~\cite{Goldberger:2004jt,Goldberger:2016iau,Damour:2016gwp,Luna:2016due,Luna:2017dtq,Bjerrum-Bohr:2018xdl,Cheung:2018wkq,Kosower:2018adc,Cristofoli:2019neg,Bjerrum-Bohr:2019kec,Mogull:2020sak,Damgaard:2021ipf, Caron-Huot:2023vxl}. Over recent years, this has culminated in state-of-the-art results at third post-Minkowski (3PM) and fourth post-Minkowski (4PM) orders.
These include the 3PM scattering angle~\cite{Bern:2019nnu,Bern:2019crd,Damour:2019lcq,Bern:2020gjj,Kalin:2020fhe}, emitted energy~\cite{Parra-Martinez:2020dzs,Damour:2020tta,Herrmann:2021lqe,Mougiakakos:2021ckm,Herrmann:2021tct,DiVecchia:2021bdo,Bini:2021gat,DiVecchia:2022nna,Kalin:2022hph} and angular momentum~\cite{Manohar:2022dea,DiVecchia:2022owy,DiVecchia:2022piu} for spinless compact objects; the 4PM conservative scattering angle~\cite{Bern:2021dqo,Dlapa:2021npj,Bern:2021yeh,Dlapa:2021vgp} and radiative effects~\cite{Dlapa:2022lmu,Dlapa:2023hsl,Damgaard:2023ttc}; results for spin effects~\cite{Bern:2020buy,Kosmopoulos:2021zoq,Liu:2021zxr,Aoude:2021oqj,Jakobsen:2021zvh,Haddad:2021znf,Aoude:2022trd,FebresCordero:2022jts,Riva:2022fru,Heissenberg:2023uvo,Bern:2023ity,Luna:2023uwd}, tidal corrections~\cite{Bern:2020uwk,Haddad:2020que,Aoude:2020ygw,AccettulliHuber:2020dal,Mougiakakos:2022sic,Heissenberg:2022tsn,Jones:2023ugm,Jakobsen:2023pvx,Riva:2023rcm}, self-force corrections~\cite{Barack:2022pde,Barack:2023oqp,Adamo:2023cfp,Kosmopoulos:2023bwc,Cheung:2023lnj} and waveforms~\cite{Jakobsen:2021smu,Cristofoli:2021vyo,Jakobsen:2021lvp,Jakobsen:2022psy,Adamo:2022qci,Elkhidir:2023dco,Brandhuber:2023hhy,Herderschee:2023fxh,Georgoudis:2023lgf,DeAngelis:2023lvf,Bini:2023fiz,Brandhuber:2023hhl,Aoude:2023dui,Georgoudis:2023eke,Georgoudis:2023ozp,Bohnenblust:2023qmy,Bini:2024rsy}. Radiative effects appear at 3PM and higher orders, and a complete account of the 3PM radiation reaction requires integrating over exchanged gravitons in the entire soft region \cite{Parra-Martinez:2020dzs, DiVecchia:2021ndb, Kalin:2022hph}. The exchanged gravitons can be considered in terms of their potential and soft contributions. While potential modes contribute to the conservative part of the scattering angle, the soft graviton modes are required for the complete 3PM radiation reaction result.

Gravitational wave observables follow from classical limits of scattering amplitudes. In this regard, there are on-shell amplitude~\cite{Bern:2019crd} and effective field theory~\cite{Dlapa:2023hsl} formalisms, the worldline formalism~\cite{Mogull:2020sak}, the KMOC formalism~\cite{Kosower:2018adc} and the eikonal approximation~\cite{DiVecchia:2023frv} to name a few of the consistent approaches for deriving these observables.  The eikonal approximation is based on a suitable Fourier transform of scattering amplitudes to impact parameter space and involves a resummation over graviton exchanges. This manifests in an eikonal phase with an all-loop order expansion, from which classical PM observables can be derived. 
While the eikonal phase is real up to tree and one-loop amplitudes, it has a pure imaginary contribution beginning at two-loops, a consequence of radiative (and, in general, inelastic) effects. This has motivated a generalized eikonal operator ansatz wherein eikonal amplitudes involving inelastic exchanges can be described as elastic eikonal amplitudes with a coherent graviton dressing operator~\cite{Cristofoli:2021jas, Britto:2021pud, DiVecchia:2022owy, DiVecchia:2022nna}. The infrared divergent contribution in the imaginary part of the eikonal is relevant for the 3PM result in that it is equivalent to the 3PM radiation reaction contribution from the real part of the eikonal and the scattering angle~\cite{DiVecchia:2021ndb,DiVecchia:2021bdo,Herrmann:2021tct,Damour:2020tta,Heissenberg:2021tzo,DiVecchia:2023frv}.
Additionally, the infrared divergent contribution is simply the Weinberg soft graviton factor and has led to a conjectured relationship between soft factors and radiation reaction~\cite{DiVecchia:2021ndb}. 
A coherent dressing constructed from the Weinberg soft graviton factor was considered in \cite{DiVecchia:2022owy, DiVecchia:2022nna} and recovers the $\mathcal{O}(G)$ expression for the leading memory effect, as well as contributions to the 3PM energy spectrum and angular momentum in the $\omega \to 0$ limit. The 3PM static contributions result from the $\mathcal{O}(G)$ expressions by carrying out the sum over external particles in terms of the impulse up to 2PM order~\cite{DiVecchia:2022piu}.

Inelastic eikonal amplitudes at three loops and higher have thus far not been derived. The gravitational dressing to higher loops can be expected to involve generalizations of the coherent dressing to multiple graviton modes and would be required for 4PM and higher gravitational wave observables. In this paper, we explore the eikonal operator ansatz up to double graviton corrections of the Weinberg soft graviton factor by utilizing the soft factorization of eikonal amplitudes. The Weinberg soft graviton factor is universal for all Lorentz invariant amplitudes and thereby provides the unique leading soft contribution from a single graviton dressing \cite{Weinberg:1965nx}. However, apart from the first subleading single soft graviton factor~\cite{Cachazo:2014fwa}, more subleading soft factors and multiple soft particle contributions are theory-specific and involve loop corrections. Hence, the generalization of the single graviton Weinberg dressing in~\cite{DiVecchia:2022owy, DiVecchia:2022nna} to multiple gravitons must be specifically derived from the amplitudes under consideration, which in our case are for eikonal amplitudes. 
In this context, we note that the generalized Wilson line (GWL) approach, based on the Schwinger formalism for propagators, derives soft factors for eikonal amplitudes by taking into account corrections to the eikonal trajectories of hard external particles~\cite{White:2011yy,Bonocore:2021qxh}.  Besides the eikonal Weinberg single soft graviton factor, the GWL also involve `next-to-eikonal' (NE) corrections from subleading multiple soft graviton contributions, which result from integrating over fluctuations about the straight-line trajectories of external particles in the amplitude. The classical universal terms in the corrected soft factor can be identified in the $\hbar \to 0$ limit. As such, the leading NE correction provides the classical limit to soft factors involving a double graviton vertex for eikonal amplitudes, which can be considered as a subleading correction to the single vertex Weinberg soft graviton factor. 
We also note that worldline techniques have recently led to a worldline quantum field theory approach to derive gravitational wave observables from gravitationally interacting binary systems~\cite{Mogull:2020sak, Jakobsen:2021smu, Jakobsen:2022psy}. This approach also involves a similar integration over fluctuations of the asymptotic straight-line path trajectories of the external particles. The worldline quantum field theory and GWL approaches agree in their classical limits for spinless external states~\cite{Bonocore:2021qxh}.

We consider the GWL by including the leading NE double graviton correction to find a real-graviton dressing operator similar to squeezed coherent states~\cite{GerryKinght} but with continuous frequency modes. In addition, the dressing a priori is not manifestly gauge invariant but can be made so by requiring the two gravitons therein to be collinear. We accordingly find a collinear double graviton dressing that can be expressed as a product of two exponential operators. One is quadratic in graviton modes, while the other is a coherent dressing operator containing corrections to all odd powers in $\kappa = \sqrt{8 \pi G}$, with $G$ Newton's constant. 
We use this dressing to define eikonal amplitudes involving inelastic exchanges up to NE collinear double graviton emissions and establish that only the coherent term contributes to classical gravitational wave observables. Due to involving corrections to all odd powers in $\kappa$, the coherent term in the dressing can be used to derive $\mathcal{O}(G^2)$ (2PM and higher PM) gravitational wave observables.

We assume this dressing provides the imaginary part of the eikonal phase in inelastic eikonal amplitudes. This allows us to derive $\mathcal{O}(G^2)$ observables from the $\kappa^3$ collinear double graviton correction in the coherent dressing. The observables we consider are the waveform, emitted momentum spectrum, and angular momentum, which are derived from expectation values with respect to corresponding soft graviton dressing states. As in the case of observables derived using the Weinberg soft factor dressing in~\cite{DiVecchia:2022owy, DiVecchia:2022nna}, our results are sensitive to the $i0$ prescription in the soft factor poles for each graviton. We find $\mathcal{O}(G^2)$ results for the memory, emitted momentum spectrum and angular momentum, that depends linearly on an upper cut-off ($\omega$) on the sum over the frequencies (energies) of the two gravitons. While these results correspond to results in a leading soft expansion, they are not `static' in that they vanish in the strict $\omega \to 0$ limit.  We also demonstrate that in an appropriate ultrarelativistic limit of the external particles, the $\mathcal{O}(G^2)$ memory approximates the nonlinear memory effect following the emission of a nearly soft graviton with frequency $\omega$. This provides a consistency check on the double graviton vertex representing the emission of a graviton from a particle following its recoil from a preceding emission.

The organization of our paper is as follows. In Sec. 2, we review the GWL formalism to derive a double graviton dressing from NE corrections of external particles in eikonal amplitudes. We substitute real graviton modes and define a manifestly gauge invariant collinear double graviton dressing. Using the Baker-Campbell-Hausdorff (BCH) formula, the dressing is expressed as a product of exponential operators that are linear and quadratic in graviton modes, respectively. The coherent contribution, i.e., the exponential operator involving one graviton mode, is shown to involve all odd powers in $\kappa$ corrections of the Weinberg soft factor. In Sec. 3, we consider the dressing as the imaginary contribution in the eikonal phase of inelastic eikonal amplitudes and discuss the expectation values of graviton mode operators with respect to the corresponding dressing states. Non-vanishing classical limits of expectation values are shown to result only from the coherent operator, with $\mathcal{O}(G^2)$ gravitational wave observables resulting from $\mathcal{O}(\kappa^3)$ contributions in the coherent term. In Sec. 4, we derive results for the waveform, emitted momentum spectrum, and angular momentum. We conclude with a discussion on future directions. Some technical details about the general form of the NE gravitational dressing factor and the kinematic form factors for the gravitational wave observables are provided in the Appendices.

\section{Double graviton dressing}

We are interested in double graviton contributions to soft factors consistent with the eikonal approach in a gravitationally interacting scalar field theory. Based on the worldline approach for propagators of external particles, we can derive the soft graviton dressing factor as the GWL from a soft expansion of the background gravitational field. The formalism generally provides a dressed propagator with subleading corrections of double and higher graviton vertices to the leading, universal Weinberg soft factor.

In this section, we first review the GWL derivation and essential properties of the dressed propagator. We will then substitute real graviton modes in the dressing operator to derive a subleading double graviton dressing for eikonal scattering processes. Unlike the single graviton Weinberg soft factor, we demonstrate that the double graviton vertex contribution in the dressing is not gauge invariant. We remedy this by considering a collinear limit to construct a gauge invariant double graviton dressing relevant for gravitational wave observables.

\subsection{Eikonal dressing from generalized Wilson line}

In this subsection, we will review the soft dressing for propagators arising from a worldline approach of field theories minimally coupled to gravity following~\cite{White:2011yy, Bonocore:2021qxh}, which we refer to for further details. The approach is based on the Schwinger proper time formalism, which expresses propagators in quantum field theory as path integrals in quantum mechanics. More specifically, the propagator can be expressed in terms of integrals over the proper time of the trajectory of the external particle. The background is considered in a PM expansion, while fluctuations about the particle trajectories are considered about asymptotic eikonal trajectories. This leads to the derivation of a dressed propagator, with the dressing factor containing multiple vertex contributions from the emission of soft gravitons. 

As in~\cite{White:2011yy, Bonocore:2021qxh} we consider a minimally coupled massive scalar field coupled to gravity
\begin{equation}
S = -\int d^4x \sqrt{-g} \left(g^{\mu \nu} \partial_{\mu}\phi^* \partial_{\nu}\phi + m^2 \phi^* \phi \right)\,.
\label{act.sf}
\end{equation}
Introducing a weak field PM expansion of the background metric about flat spacetime
\begin{equation}
g_{\mu \nu} = \eta_{\mu \nu} + 2 \kappa h_{\mu \nu}\,,
\label{met}
\end{equation}
with $\kappa^2 = 8 \pi G$, and expanding \ref{act.sf} up to second order in $h_{\mu \nu}$ and its inverse, we find an action of the form
\begin{align}
S =- \int d^4x\, \phi^* \left(2 H \right) \phi\,  
\label{SH}
\end{align}
with
\begin{align}
2 H(x\,, p) & = p^2 + m^2 + 2 \kappa \left[ -p^{\mu} p^{\nu}h_{\mu \nu} + i p^{\nu} \partial^{\mu} h_{\mu \nu} + \frac{1}{2}\left(p^2 + m^2\right) h - \frac{i}{2}p^{\mu} \partial_{\mu} h  \right] \notag\\
&  \qquad + 4 \kappa^2 \left[\left(p^2 + m^2\right) \left(\frac{1}{8}h^2 - \frac{1}{4} h_{\mu \nu} h^{\mu \nu} \right) - i p^{\mu} \partial_{\mu}\left(\frac{1}{8}h^2 - \frac{1}{4}h_{\mu \nu} h^{\mu \nu} \right)  \right.  \notag\\
& \left. \quad \qquad \qquad  + p^{\mu}p^{\nu}\left(h_{\mu \rho} h_{\nu}^{\rho} - \frac{1}{2}h_{\mu \nu} h\right) - i p^{\mu}\partial^{\nu} \left(h_{\mu \rho} h_{\nu}^{\rho} - \frac{1}{2} h_{\mu \nu} h\right)\right] + \mathcal{O}(\kappa^3)\,,
\label{H.exp}
\end{align}
and where all indices are contracted with the flat spacetime metric. The expression in \ref{SH} results from replacing $\partial_{\mu}$ on the scalar field with $i p_{\mu}$, and the $px$-ordering to keep all momenta to the left\footnote{This is not the same replacement as in \cite{Bonocore:2021qxh}. We have the most plus signature for the metric, with the action as in \ref{act.sf}. We get the same result as in \cite{Bonocore:2021qxh} with $p_{\mu} = -i \partial_{\mu}$ instead of $p_{\mu} = i \partial_{\mu}$ when turning $H$ into Hamiltonian operator $\hat{H}$ in \ref{prop.sch} of Schwinger formalism.}. Hence turning $H$ into the Hamiltonian operator $\hat{H}$, we note that it governs the evolution of a scalar particle in the background field.
This interpretation is made manifest by expressing the dressed scalar propagator $(\hat{H}-i 0)^{-1}$ (with chosen $i 0$ prescription for causality) in the Schwinger representation as a double path integral over position and momentum space with initial position $x_i$ and final momentum $p_f$, 
\begin{align}
\Big \langle p_f \Big \vert &\left(\hat{H} - i 0\right)^{-1} \Big\vert x_i \Big \rangle \notag\\
& = \int_{0}^{\infty} dT \int_{x(0) = x_i}^{p(T) = p_f} \mathcal{D}p \mathcal{D}x \;\exp\left[ - \frac{i}{\hbar}p(T)\cdot x(T) +\frac{i}{\hbar} \int_{0}^T dt \left( p\cdot \dot{x} - \hat{H}(x\,, p) + i0 \right) \right]\,,
\label{prop.sch}
\end{align}
where $T$ is the Schwinger parameter.

To derive the soft graviton dressing within the eikonal approximation, one considers the worldline of a freely propagating scalar particle subject to small fluctuations arising from background soft gravitons, i.e. 
\begin{equation}
x(t) = x_i + p_f t + \tilde{x}(t) \;; \quad p(t) = p_f + \tilde{p}(t) \,,
\label{xp.sol}
\end{equation}
with $\tilde{x}(t)$ and $\tilde{p}(t)$ representing perturbations about the straight-line classical trajectory of the particle, which are subject to the boundary conditions $\tilde{p}(T) = 0 = \tilde{x}(0)$.  The case with $x_i = 0$ in \ref{xp.sol} provides a convenient choice to evaluate~\ref{prop.sch}, while $x_i \neq 0$ will provide orbital angular momentum contributions of the external particle~\cite{White:2011yy, Bonocore:2021qxh,Luna:2016idw}. In the following, we will adopt the $x_i = 0$ choice for the external particle trajectories. 
As the unperturbed trajectory results in a strict soft graviton exchange limit, the perturbations allow for a systematic derivation of the eikonal limit and sub-eikonal corrections. The latter result from the external particle recoil due to emitted gravitons and, in general, produce contributions corresponding to multiple graviton vertices. To quadratic order in $h_{\mu \nu}$, we have the leading next-to-eikonal (NE) correction that provides a double graviton emission vertex contribution to the soft graviton dressing. This is derived as a generalized Wilson line (GWL) from the amputated dressed scalar propagator \cite{Laenen:2008gt, Bonocore:2021qxh} 
\be
\lim_{T\to \infty} \widetilde{W}_{p_f}(0,T) := \lim_{p_f^2 \to -m^2}(p_f^2+m^2-i 0) \Big \langle p_f \Big \vert \left(\hat{H} - i 0\right)^{-1} \Big\vert x_i \Big \rangle \;. 
\label{sf.d}
\ee
Hence $\displaystyle{\widetilde{W}_{p_f}(0,\infty) := \lim_{T \to \infty} \widetilde{W}_{p_f}(0,T)}$ is the soft dressing of an asymptotic state. The expression in \ref{sf.d} requires integrating over $\tilde{p}$ and $\tilde{x}$. The $\tilde{p}$ integral is Gaussian, while the integral over $\tilde{x}$ requires the use of correlators
\begin{align}
\langle x_{\mu} (t) x_{\nu}(t') \rangle = i \text{min} (t,t')\eta_{\mu \nu} \;; \quad \langle \dot{x}_{\mu} (t) x_{\nu}(t') \rangle = i \Theta(t'-t)\eta_{\mu \nu} \;;\quad  \langle \dot{x}_{\mu} (t) \dot{x}_{\nu}(t') \rangle =& i \delta(t -t')\eta_{\mu \nu} 
\label{x.con}
\end{align}
which follow from the Green's function for $x_{\mu}(t)$. The evaluation of \ref{sf.d} provides the dressing 
\be 
\widetilde{W}_{p_f}\left(0\,, \infty\right) := e^{-\Delta_{p_f}}
\ee
with 
\begin{align}
\Delta_{p_f} &= - \frac{i \kappa}{\hbar} \int_{0}^{\infty} dt \left[ -p^{\mu}_f p^{\nu}_f  + i p^{(\mu}_f \partial^{\nu)} - \frac{i}{2}\eta^{\mu \nu} p^{\alpha}_f  \partial_{\alpha} + \frac{i}{2} t p^{\mu}_f p^{\nu}_f \partial^2 \right] h_{\mu \nu}(p_f t) \notag\\
&  \quad \qquad + \frac{2 i \kappa^2}{\hbar} \int_{0}^{\infty} dt\int_{0}^{\infty} ds \left[\frac{p^{\mu}_f p^{\nu}_f p^{\rho}_f p^{\sigma}_f}{4} \text{min}(t,s) \partial_{\alpha}h_{\mu \nu} (p_f t) \partial^{\alpha}h_{\rho \sigma} (p_f s) \right. \notag\\
& \left. \phantom{\frac{p^{\mu}_f}{4}}\quad \quad + p^{\mu}_f p^{\nu}_f p^{\rho}_f \Theta(t\,,s) h_{\rho \sigma} (p_f s)\partial^{\sigma}h_{\mu \nu} (p_f t) + p^{\nu}_f p^{\rho}_f \delta(t - s) \eta^{\mu \sigma}h_{\rho \sigma} (p_f s) h_{\mu \nu} (p_f t) \right] \;.
\label{gwl.drs2}
\end{align}

The above discussion for deriving GWL for a scalar particle is subtly different from the worldline quantum field theory formalism \cite{Mogull:2020sak, Jakobsen:2021smu, Jakobsen:2022psy}, in which the dressed propagator is utilized as an efficient tool in reorganizing the Feynman rules for evaluating the scattering amplitudes.  On the other hand, the GWL of \ref{gwl.drs2} encodes the collective effect of the background field as a soft dressing associated with the external particles in an amplitude. Consider the scattering of $N$ particles with asymptotic momenta 
$p_n^{\mu} = \eta_n \left(E_n \,, \vec{p}_n\right)$, where $\eta=+1$ for outgoing particles and $\eta = -1$ for incoming particles in the all-outgoing convention. Then the overall soft graviton dressing operator denoted by $e^{-\Delta}$ 
is just the product of GWL dressings on each external particle and is given by 
\begin{align}
\exp [-\Delta ]=\exp \left[-\sum_{n=1}^N \Delta_{p_n} \right]\;.
\label{gwl.drs}
\end{align} 

We will next consider real graviton modes in the eikonal dressing operator and study its properties.

\subsection{Eikonal dressing operator with real graviton modes}

We consider two graviton modes with their wavenumbers parametrized by
\begin{align}
k^{\mu} &= \omega_k q^{\mu}_k \;  \;; \qquad  q^{\mu}_k = (1\,,\hat{k}) \;,\notag\\
l^{\mu} &= \omega_l q^{\mu}_l \; \;;  \qquad q^{\mu}_l = (1\,,\hat{l}) \;, \label{mom.param}
\end{align}
with $\omega_k$ and $\omega_l$ denoting their frequencies, $\hat{k}$ and $\hat{l}$ their spatial unit normal vectors indicating orientations, and $q^2_{k} = 0= q^2_l$. The graviton polarization tensors will be denoted by $\varepsilon_{i\,,\mu \nu} (k)$ and $\varepsilon_{j\,,\rho \sigma} (l)$. We then have the following conventional definition for two real graviton modes
\begin{align}
h_{\mu \nu} (x) &= \int_{\vec{k}} d^3k \left[ a_i (k) \varepsilon_{i\,,\mu \nu} (k) e^{ik\cdot x} + a^{\dagger}_i (k) \varepsilon^*_{i\,,\mu \nu} (k) e^{-ik\cdot x} \right] \label{gm}\,,\\
h_{\rho \sigma} (x') &= \int_{\vec{l}} d^3l \left[ a_j (l) \varepsilon_{j\,,\rho \sigma} (l) e^{il\cdot x'} + a^{\dagger}_j (l) \varepsilon^*_{j\,,\rho \sigma} (l) e^{-il\cdot x'} \right] \;,\label{gm.l}
\end{align}
where repeated Latin indices without the explicit summation symbol denotes a sum over polarizations. The on-shell integration measure explicitly has the form
\begin{equation}
\int_{\vec{k}} d^3k \equiv \int
\frac{d^3k}{(2 \pi)^3 2 \omega_k} = \frac{1}{2 (2 \pi)^3} \int_{0}^{\infty} \Theta\left(\omega - \omega_k \right) d \omega_k \omega_k \oint d\Omega_k\,.
\label{meas}
\end{equation}
The indicated $\omega_k$ integral in \ref{meas} is over a soft region with an upper cut-off frequency $\omega$ as the maximal resolution of the detector in the sense of Bloch and Nordsieck for dealing with IR divergences \cite{Bloch:1937pw}. When we integrate over several real gravitons, the Heaviside step function argument gets replaced by one with the sum over all $\omega_k$ being less than $\omega$. 

Following the conventions of~\cite{DiVecchia:2022nna,DiVecchia:2022owy,DiVecchia:2022piu} \footnote{This choice introduces an overall $\hbar$ in \ref{del.def} in an otherwise wavenumber convention, which will be relevant in this paper for the derivation of classical expectation values from the gravitational dressing. We can go to a momentum description to find expressions with all $\hbar$ dependencies manifest. In the adopted conventions, by replacing $\mathbf{k}^{\mu} = \hbar k^{\mu}$, $a_{i}(\mathbf{k}) = \hbar^{-\frac{3}{2}} a_{i}(k)$ and $a^{\dagger}_{i}(\mathbf{k}) = \hbar^{-\frac{3}{2}} a_{i}(k)$, where $\mathbf{k}^{\mu}$ is the corresponding graviton momentum (and similarly $\mathbf{l}^{\mu}$ for the other graviton), we recover momentum expressions consistent with~\cite{Kosower:2018adc,Cristofoli:2021vyo,Cristofoli:2021jas}. The graviton polarization tensors are the same in wavenumber and momentum descriptions.}, the graviton annihilation and creation operators, respectively $a_i$ and $a^{\dagger}_i$ in \ref{gm} and \ref{gm.l}, satisfy the commutation relation
\begin{equation}
\left[a_i(k) \,, a_j^{\dagger}(k')\right] = \delta(\vec{k}\,, \vec{k}') \delta_{ij} \,,  \label{mode.comm}
\end{equation}
with $\delta_{ij}$ the Kronecker delta for the polarization indices and $\delta(\vec{k}\,, \vec{k}')$ the Dirac delta function defined with respect to the on-shell measure for real gravitons 
\begin{equation}
\delta(\vec{k}\,, \vec{k}') = 2 \hbar \omega_{k} (2 \pi)^3 \delta^{(3)}(\vec{k} - \vec{k}') \;. \label{del.def}
\end{equation}

To evaluate \ref{gwl.drs}, we consider two soft graviton dressing modes by setting $x^{\mu} = p_n^{\mu} t$ and $x'^{\mu} = p_n s$ in \ref{gm} and \ref{gm.l} respectively. This gives
\begin{align}
h_{\mu \nu} (p_nt) &= \int_{\vec{k}} d^3k \left[ a_i (k) \varepsilon_{i\,,\mu \nu} (k) e^{ik\cdot p_n t} + a^{\dagger}_i (k) \varepsilon^*_{i\,,\mu \nu} (k) e^{-ik\cdot p_n t} \right] \;, \notag\\
h_{\rho \sigma} (p_ns) &= \int_{\vec{l}} d^3l \left[ a_j (l) \varepsilon_{j\,,\rho \sigma} (l) e^{il\cdot p_n s} + a^{\dagger}_j (l) \varepsilon^*_{j\,,\rho \sigma} (l) e^{-il\cdot p_n s} \right] \;. \label{g2m}
\end{align}

Substituting \ref{g2m} in \ref{gwl.drs}, the integrals over time are evaluated using the relations
\begin{align}
\int_0^{\infty} dt \int_0^{\infty} ds\; \text{min}(t,s) \exp[i a t] \exp [i b s] = \frac{-i}{ a\, b \, (a+b)} \;,\label{min.int} \\
\int_0^{\infty} dt \int_0^{\infty} ds\; \Theta(t - s) \exp[i a t] \exp [i b s] =  \frac{- 1}{ a \, (a+b)} \;,\label{theta.int1} \\
\int_0^{\infty} dt \int_0^{\infty} ds\; \Theta(s - t) \exp[i a t] \exp [i b s] =  \frac{- 1}{ b \, (a+b)} \;, \label{theta.int2} \\
\int_0^{\infty} dt \int_0^{\infty} ds\; \delta(t - s) \exp[i a t] \exp [i b s] = \frac{-i}{(a+b)}\;. \label{del.int}
\end{align}
This then yields 
\begin{align}
\exp \left[-\Delta\right] &= \exp\left[-\tilde{\Delta}_1 -\tilde{\Delta}_2\right] \label{dress.t}\\
\tilde{\Delta}_1 &= \frac{1}{\hbar} \int_{\vec{k}} d^3k \left(a_i (k) f^*_i(k) - a^{\dagger}_i (k) f_i(k) \right)\;, \label{d1.f}\\
\tilde{\Delta}_2 &= \frac{1}{2 \hbar}\int_{\vec{k}} d^3k \int_{\vec{l}} d^3l \left[a^{\dagger}_i (k) a^{\dagger}_j (l) \tilde{A}_{ij}(k,l) - a_i (k) a_j (l) \tilde{A}^*_{ij}(k,l)  \right. \notag\\
& \left. \qquad \qquad  \qquad \qquad \qquad + a^{\dagger}_i (k) a_j (l) \tilde{B}^*_{ij}(k,l) - a^{\dagger}_j (l) a_i (k) \tilde{B}_{ij}(k,l)\right] \label{d2.ft}
\end{align}
with the factors $f_i(k)$, $\tilde{A}_{ij}(k,l)$ and $\tilde{B}_{ij}(k,l)$ defined by
\begin{align}
f_i(k) &= \varepsilon^*_{i\,,\mu \nu} (k) F^{\mu \nu}(k) \;; \qquad \qquad F^{\mu \nu}(k)  =  \kappa \; \sum_{n} \frac{ p_n^{\mu} p_n^{\nu}}{k\cdot p_n} \label{f.def}\\
\tilde{A}_{ij}(k,l) &= \varepsilon^*_{i\,,\mu \nu} (k) \tilde{A}^{\mu\nu\rho \sigma} \varepsilon^*_{j\,,\rho \sigma} (l) \;; \notag\\
\tilde{A}^{\mu\nu\rho \sigma}(k,l) & =\kappa^2 \sum_{n} \frac{1}{p_n\cdot (k+l)}\left[ - k\cdot l \left(\frac{p_n^{\mu} p_n^{\nu}}{k\cdot p_n}\right)\left(\frac{p_n^{\rho} p_n^{\sigma}}{l\cdot p_n}\right) + \left(\frac{p_n^{\mu} p_n^{\nu}}{k\cdot p_n}\right)(p_n^{\rho} k^{\sigma} + p_n^{\sigma} k^{\rho})  \right. \notag\\
& \left. \quad  +  \left(\frac{p_n^{\rho} p_n^{\sigma}}{l\cdot p_n}\right)(p_n^{\mu} l^{\nu} + p_n^{\nu} l^{\mu}) - (p_n^{\rho} p_n^{\mu} \eta^{\nu \sigma} + p_n^{\rho} p_n^{\nu} \eta^{\mu \sigma} + p_n^{\sigma} p_n^{\mu} \eta^{\nu \rho} + p_n^{\sigma} p_n^{\nu} \eta^{\mu \rho}) \right]\;. \label{A.d1}\\
\tilde{B}_{ij}(k,l) &= \varepsilon_{i\,,\mu \nu} (k) \tilde{B}^{\mu\nu\rho \sigma} \varepsilon^*_{j\,,\rho \sigma} (l) \;; \notag\\
\tilde{B}^{\mu\nu\rho \sigma}(k,l) & =\kappa^2 \sum_{n} \frac{1}{p_n\cdot (k-l)}\left[ - k\cdot l \left(\frac{p_n^{\mu} p_n^{\nu}}{k\cdot p_n}\right)\left(\frac{p_n^{\rho} p_n^{\sigma}}{l\cdot p_n}\right) + \left(\frac{p_n^{\mu} p_n^{\nu}}{k\cdot p_n}\right)(p_n^{\rho} k^{\sigma} + p_n^{\sigma} k^{\rho}) \right. \notag\\
& \left. \quad +  \left(\frac{p_n^{\rho} p_n^{\sigma}}{l\cdot p_n}\right)(p_n^{\mu} l^{\nu} + p_n^{\nu} l^{\mu}) - (p_n^{\rho} p_n^{\mu} \eta^{\nu \sigma} + p_n^{\rho} p_n^{\nu} \eta^{\mu \sigma} + p_n^{\sigma} p_n^{\mu} \eta^{\nu \rho} + p_n^{\sigma} p_n^{\nu} \eta^{\mu \rho}) \right]\;. \label{B.d1}
\end{align}

The leading contribution to the dressing from \ref{d1.f} is the Weinberg soft graviton dressing considered in \cite{Weinberg:1965nx, DiVecchia:2022nna, DiVecchia:2022piu}. The contribution from \ref{d2.ft} is the double graviton contribution to the soft factor derived from the GWL approach to eikonal scattering. 

In arriving at \ref{dress.t}, we have omitted a vanishing exponential factor denoted as $\Delta_{\text{Remainder}}$, which is formally given by 
\begin{align}
\Delta_{\text{Remainder}} &= - \frac{1}{2}\int_{\vec{k}} d^3k \int_{\vec{l}} d^3l\; \delta(\vec{k}\,, \vec{l})\; \delta_{ij}\left(m_{ij}(k,l) - n_{ij}(k,l)\right) \,,
\label{d.rem}
\end{align}
with $m_{ij}(k,l)$ and $n_{ij}(k,l)$ defined by
\begin{align}
m_{ij}(k,l) &=  \varepsilon_{i\,,\mu \nu} (k) M^{\mu \nu \rho \sigma} (k\,,l) \varepsilon^*_{j\,,\rho \sigma} (l)\;; \notag\\
M^{\mu \nu \rho \sigma} (k\,,l)& =  \frac{1}{2} \sum_{n} \frac{\kappa^2}{p_n\cdot (k-l)} \left[-k\cdot l \left(\frac{p_n^{\mu} p_n^{\nu}}{k\cdot p_n}\right)\left(\frac{p_n^{\rho} p_n^{\sigma}}{l\cdot p_n}\right) +  2 \left(\frac{p_n^{\mu} p_n^{\nu}}{k\cdot p_n}\right)(p_n^{\rho} k^{\sigma} + p_n^{\sigma} k^{\rho})\right. \notag\\
& \left. \phantom{\left(\frac{p_n^{\rho} p_n^{\sigma}}{l\cdot p_n}\right)} \qquad\qquad \qquad \qquad   - \,   (p_n^{\rho} p_n^{\mu} \eta^{\nu \sigma} + p_n^{\rho} p_n^{\nu} \eta^{\mu \sigma} + p_n^{\sigma} p_n^{\mu} \eta^{\nu \rho}+ p_n^{\sigma} p_n^{\nu} \eta^{\mu \rho}) \right]\;. \label{m.def}\\
n_{ij}(k,l) &=  \varepsilon^*_{i\,,\mu \nu} (k) N^{\mu \nu \rho \sigma} (k\,,l) \varepsilon_{j\,,\rho \sigma} (l) \;; \notag\\
N^{\mu \nu \rho \sigma} (k\,,l)& = \frac{1}{2}\sum_{n} \frac{\kappa^2}{p_n\cdot (k-l)} \left[-k.l \left(\frac{p_n^{\mu} p_n^{\nu}}{k\cdot p_n}\right)\left(\frac{p_n^{\rho} p_n^{\sigma}}{l\cdot p_n}\right) + 2 \left(\frac{p_n^{\rho} p_n^{\sigma}}{l\cdot p_n}\right)(p_n^{\mu} l^{\nu} + p_n^{\nu} l^{\mu}) \right.\notag\\ 
& \left. \phantom{\left(\frac{p_n^{\rho} p_n^{\sigma}}{l\cdot p_n}\right)} \qquad \qquad \qquad \qquad   - \,  (p_n^{\rho} p_n^{\mu} \eta^{\nu \sigma} + p_n^{\rho} p_n^{\nu} \eta^{\mu \sigma} + p_n^{\sigma} p_n^{\mu} \eta^{\nu \rho}+ p_n^{\sigma} p_n^{\nu} \eta^{\mu \rho}) \right] \;.
\label{n.def}
\end{align}
Upon evaluating the integral \ref{d.rem} by utilizing the 
delta function $\delta(\vec{k}\,, \vec{l})$, the contributions from $m_{ij}(k,l)$ and $n_{ij}(k,l)$ cancel out leading to a vanishing result.

Hence, the dressing derived from the worldline approach is simply given by \ref{dress.t}. It is a unitary operator in that
\begin{equation}
\left(\exp\left[-\tilde{\Delta}_1 - \tilde{\Delta}_2\right]\right)^{\dagger} = \exp\left[\tilde{\Delta}_1 + \tilde{\Delta}_2\right]\;.
\end{equation}

However, unlike the Weinberg dressing factor $\tilde{\Delta}_1$ of \ref{d1.f}, we will see that the double graviton dressing factor $\tilde{\Delta}_2$ of \ref{d2.ft} is not gauge invariant. Next, we will address this issue to derive the relevant double graviton dressing of the eikonal process.

\subsection{Gauge invariant dressing from collinear limit}

An important property of the Weinberg soft factor is its gauge invariance when physical observables are concerned. In the following, we will discuss this property and consider it for the double graviton contribution. To this end, we implement the gauge transformation of the graviton modes of \ref{gm} and \ref{gm.l} by the following transformation on their polarization tensors, respectively, 
\begin{align}
\varepsilon_i^{\mu \nu} (k) &\longrightarrow \varepsilon_i^{\mu \nu} (k) + \xi^{\mu} k^{\nu} + \xi^{\nu} k^{\mu} = \varepsilon_i^{\mu \nu} (k) + \omega_k \left(\xi^{\mu} q_{k}^{\nu} + \xi^{\nu} q_{k}^{\mu}\right)\;, 
\notag\\
\varepsilon_i^{\rho \sigma} (l) &\longrightarrow \varepsilon_i^{\rho \sigma} (l) + \zeta^{\rho} l^{\sigma} + \zeta^{\sigma} l^{\rho} = \varepsilon_i^{\rho \sigma} (l) + \omega_l \left(\zeta^{\sigma} q_{l}^{\rho} + \zeta^{\rho} q_{l}^{\sigma}\right)\;, 
\label{gt.pol}
\end{align}
where $k^{\mu}$ and $l^{\sigma}$ takes the form of \ref{mom.param}, and $\xi^{\mu}$ and $\zeta^{\sigma}$ are reference vectors required to satisfy $\xi\cdot q_{k} = 0$ and $\zeta\cdot q_{l} = 0$. 
This condition ensures that the transverse and traceless properties of the polarization tensors are respected. 

It is straightforward to see that the gauge invariance of Weinberg soft dressing factor \ref{f.def} follows from the momentum conservation of the external hard particles 
\begin{equation}
k_{\mu} F^{\mu \nu}(k)  = \kappa \sum_{n} p_n^{\mu} = 0\;.
\label{F.tr}
\end{equation} 
Hence, the single graviton contribution to the dressing from \ref{dress.t} is gauge invariant. 

We may similarly consider gauge transformations of the double graviton contributions from \ref{A.d1} and \ref{B.d1}. The resulting gauge transformations for $\tilde{A}_{ij}(k,l)$ and $\tilde{B}_{ij}(k,l)$ can be investigated by contracting $\tilde{A}^{\mu\nu\rho \sigma}(k,l)$ and $\tilde{B}^{\mu\nu\rho \sigma}(k,l)$ with $k_{\mu} \xi_{\nu}$ for the first two indices and with $l_{\rho} \zeta_{\sigma}$ for the last two indices. We find
\begin{align}
k_{\mu} \xi_{\nu} \tilde{A}^{\mu\nu\rho \sigma}(k,l) &= \kappa^2 \sum_{n} \frac{p_n\cdot k}{p_n\cdot (k+l)}\left[ \left(\frac{p_n^{\rho}p_n^{\sigma}}{p_n\cdot q_l}\right) \xi\cdot q_l - \left(p_n^{\rho} \xi^{\sigma}  + p_n^{\sigma} \xi^{\rho}\right)\right] \;, 
\notag\\
k_{\mu} \xi_{\nu} \tilde{B}^{\mu\nu\rho \sigma}(k,l) &= \kappa^2 \sum_{n} \frac{p_n\cdot k}{p_n\cdot (k-l)}\left[ \left(\frac{p_n^{\rho}p_n^{\sigma}}{p_n\cdot q_l}\right) \xi\cdot q_l - \left(p_n^{\rho} \xi^{\sigma}  + p_n^{\sigma} \xi^{\rho}\right)\right]\;, 
\notag\\
l_{\rho} \zeta_{\sigma}\tilde{A}^{\mu\nu\rho \sigma}(k,l) &= \kappa^2 \sum_{n} \frac{p_n \cdot k}{p_n\cdot (k+l)}\left[ \left(\frac{p_n^{\mu}p_n^{\nu}}{p_n\cdot q_k}\right) \zeta\cdot q_k - \left(p_n^{\mu} \zeta^{\nu}  + p_n^{\nu} \zeta^{\mu}\right)\right]\;, 
\notag\\
l_{\rho} \zeta_{\sigma} \tilde{B}^{\mu\nu\rho \sigma}(k,l)  &= \kappa^2 \sum_{n} \frac{p_n\cdot k}{p_n\cdot (k - l)}\left[ \left(\frac{p_n^{\mu}p_n^{\nu}}{p_n\cdot q_k}\right) \zeta\cdot q_k - \left(p_n^{\mu} \zeta^{\nu}  + p_n^{\nu} \zeta^{\mu}\right)\right]\;.  
\label{ab.gt}
\end{align}
 
As all the terms in \ref{ab.gt} do not vanish, the double graviton terms in \ref{A.d1} and \ref{B.d1} are not gauge invariant. One way this can be remedied, which is particularly appropriate in the context of asymptotic gravitational wave observables, is to consider a collinear limit wherein $\hat{k} = \hat{l}$. This can be formally implemented through the use of a delta function over the angular variables, $\delta(\Omega_{k}\,, \Omega_{l})$, which satisfies 
\begin{equation}
\oint d \Omega_k \delta(\Omega_{k}\,, \Omega_{l}) = 1 = \oint d \Omega_l \delta(\Omega_{k}\,, \Omega_{l}) \;.
\end{equation}

We accordingly define  
\begin{align}
&\qquad   A_{ij}(k,l) = \varepsilon^*_{i\,,\mu \nu} (k) A^{\mu\nu\rho \sigma} \varepsilon^*_{j\,,\rho \sigma} (l) \;; \qquad B_{ij}(k,l) = \varepsilon_{i\,,\mu \nu} (k) B^{\mu\nu\rho \sigma} \varepsilon^*_{j\,,\rho \sigma} (l)\;;  \notag\\
&A^{\mu\nu\rho \sigma}(k,l) =  \tilde{A}^{\mu\nu\rho \sigma}(k,l) (2 \pi)^2 \delta(\Omega_{k}\,, \Omega_{l}) \;; \qquad B^{\mu\nu\rho \sigma}(k,l) = \tilde{B}^{\mu\nu\rho \sigma}(k,l) (2 \pi)^2 \delta(\Omega_{k}\,, \Omega_{l})\,,
\label{nAB.def}
\end{align}
where $\tilde{A}^{\mu\nu\rho \sigma}(k,l)$ and $\tilde{B}^{\mu\nu\rho \sigma}(k,l)$ are as previously defined in \ref{A.d1} and \ref{B.d1}, and the factor of $(2 \pi)^2$ has been included to account for the corresponding inverse factor present in the on-shell graviton measure. The gauge transformations of $A_{ij}(k,l)$ and $B_{ij}(k,l)$ in \ref{nAB.def} can be investigated through contractions with $A^{\mu\nu\rho \sigma}(k,l)$ and $B^{\mu\nu\rho \sigma}(k,l)$, as in \ref{ab.gt}. For the first equation in \ref{ab.gt} we now find
\begin{align}
k_{\mu} \xi_{\nu} A^{\mu\nu\rho \sigma}(k,l) &=  (2 \pi)^2\delta(\Omega_{k}\,, \Omega_{l}) \frac{\kappa^2 \omega_k}{\omega_k + \omega_l}\sum_{n} \left[ \left(\frac{p_n^{\rho}p_n^{\sigma}}{p_n\cdot q_k}\right) \xi \cdot q_k - \left(p_n^{\rho} \xi^{\sigma}  + p_n^{\sigma} \xi^{\rho}\right)\right] = 0\,,
\label{abn.gt}
\end{align}
with similar transformations for the other equations in \ref{ab.gt}. The collinear limit ensured by  $\delta(\Omega_{k}\,, \Omega_{l})$ allows us to reduce $\frac{p_n\cdot k}{p_n\cdot (k\pm l)}$ to a $n$-independent form of $\frac{ \omega_k}{\omega_k \pm \omega_l} $, and to interchange $q_k$ and $q_l$ freely in \ref{ab.gt}, thus resulting in \ref{abn.gt}. Note that the first term in the parenthesis of \ref{abn.gt} vanishes due to $\xi\cdot q_k = 0 = \zeta\cdot q_l$, while the last two terms vanish by momentum conservation. Hence the collinear modification results in the gauge invariance of the double graviton soft factor under shifts of the polarization tensor.

The definition in \ref{nAB.def} is thus gauge invariant and we can now change the dressing in \ref{dress.t} to 
\begin{align}
\exp \left[-\Delta\right] &= \exp\left[-\tilde{\Delta}_1 -\Delta_2\right] \label{dress}\\
\Delta_2 &= \frac{1}{2 \hbar}\int_{\vec{k}} d^3k \int_{\vec{l}} d^3l \left[a^{\dagger}_i (k) a^{\dagger}_j (l) A_{ij}(k,l) - a_i (k) a_j (l) A^*_{ij}(k,l)  \right. \notag\\
& \left. \qquad \qquad  \qquad \qquad \qquad + a^{\dagger}_i (k) a_j (l) B^*_{ij}(k,l) - a^{\dagger}_j (l) a_i (k) B_{ij}(k,l)\right] \label{d2.f}\,,
\end{align}
with no modification to $\tilde{\Delta}_1$ in \ref{d1.f}, and $\Delta_2$ being the collinear limit of $\tilde{\Delta}_2$ of \ref{d2.ft}. Hence $\tilde{\Delta}_1$ is the Weinberg soft graviton factor, while $\Delta_2$ provides a $\kappa^2$ correction due to two collinear gravitons. We note that \ref{dress} can also be derived from the analysis in the previous subsection by considering \ref{gm.l} with a $\delta(\Omega_{k}\,, \Omega_{l})$ in the integrand to ensure collinearity.

We make some general remarks on the above in light of soft factor results in the literature. It is known that the double soft graviton factor for field theories is manifestly gauge invariant~\cite{Chakrabarti:2017ltl,Chakrabarti:2017zmh,Distler:2018rwu,Marotta:2020oob} {\footnote{We thank Paolo Di Vecchia and Carlo Heissenberg for bringing these references and the gauge invariance of the double soft graviton factor to our attention}}. From a perturbative field theory perspective, we expect gauge invariance if all terms to the same order in coupling constant are included. The terms appearing in \ref{A.d1} and \ref{B.d1} capture recoil effects on the external particles due to two soft gravitons. More specifically, these represent the seagull and Born contributions noted in~\cite{Distler:2018rwu}. In addition to these terms recovered by the GWL approach, there are contributions from two successive single graviton emissions and a double graviton pole, and these terms are required for the gauge invariance of the complete double soft graviton factor. As the GWL soft factor derivation only made use of external particle trajectories in a background gravitational field and not the dynamics of gravitons, these additional pieces are absent in the GWL result. Thus collinearity, and generally the relative orientations of the gravitons, do not appear to have their dynamical origin in the GWL approach. We find that the contributions in \ref{A.d1} and \ref{B.d1} can be rendered gauge invariant by imposing collinearity and hence restricting their relative orientations.  This reduces the double soft graviton factor contribution to that in \ref{dress}, which only involves a double graviton vertex or, equivalently, a collinear seagull contribution. The substitution of collinear gravitons in the GWL soft factor may thus be considered a natural requirement to derive a manifestly gauge invariant dressing to NE orders.

We have also not accounted for orbital angular momentum contributions, as noted through our choice of $x_i = 0$ in the parametrized eikonal trajectory of \ref{xp.sol}. In \cite{Luna:2016idw}, the GWL formalism was applied to eikonal scattering amplitudes in QCD and gravity, with the Wilson line dressing considered up to the first single soft subleading correction with a non-vanishing initial offset $x_i \neq 0$. This results in a next-to-soft corrected eikonal amplitude, resulting from internal hard contributions as well as external soft emissions. The latter exponentiates and involves single soft graviton corrections (still at order $\kappa$) of the known Weinberg soft graviton dressing. The internal contributions, however, do not exponentiate and can be determined by the action of the angular momentum operator acting on the leading eikonal amplitude. The complete next-to-leading order correction of the amplitude in the Regge limit follows from the combined contributions of internal and external emissions, and establishes angular momentum conservation associated with the subleading single soft factor. Interestingly, the internal emission graphs include those with seagull and triple graviton vertices, along with contributions from momentum shifts of the external particles in the amplitude. However, the seagull and other internal emission graphs in~\cite{Luna:2016idw} are distinct from those considered in this section in two ways. First, the next-to-leading order corrected Born amplitude in~\cite{Luna:2016idw} involves a single external graviton while \ref{dress} involves two external gravitons, with the corresponding double soft factor sensitive to the sum over their momenta. Secondly, the internal emission contributions in the GWL approach do not exponentiate, while \ref{dress} is nothing but the GWL dressing to $\kappa^2$ order in a collinear limit due to external graviton emissions. Our expression in \ref{dress} follows from \ref{gwl.drs2}, whose $\kappa^2$ double graviton corrections are precisely the quadratic in coupling constant terms not considered in the external emission contributions of~\cite{Luna:2016idw} (as their analysis was considered to one-loop order).

\section{Expectation values of radiative observables}

Following the eikonal operator approach in~\cite{DiVecchia:2022nna,DiVecchia:2022owy,DiVecchia:2022piu,DiVecchia:2023frv}, we now identify the dressing of the previous section as that for the full $S$-matrix of the hard elastic eikonal scattering process 
\be
S \approx e^{-\Delta} e^{ i \text{Re} 2 \delta}
\label{s.eik}
\ee
where the real part of the eikonal phase $\delta$ is included, in addition to the GWL soft factor $\Delta$ as its imaginary part. However, the exact form of $\text{Re} 2 \delta$ will not relevant for our consideration of gravitational wave observables. If we consider \ref{s.eik} in impact parameter space, the external particle momenta in the soft factor should be appropriately identified with derivatives with respect to the impact parameter~\cite{DiVecchia:2022nna,DiVecchia:2022owy,DiVecchia:2022piu,DiVecchia:2023frv}. We briefly explain the `$\approx$' symbol in \ref{s.eik}. The complete eikonal amplitude involves an overall factor of $\left( 1 + 2 i \Delta(\sigma,b)\right)$, where $\Delta(\sigma,b)$ is a quantum remainder. However, the quantum remainder does not contribute to classical expectation values and thus has not been explicitly considered in the above expression. Additionally, the general result follows from a Fourier transform of the eikonal amplitude from momentum space to impact parameter space. The eikonal operator can involve both soft and non-soft graviton modes, with the latter sensitive to the integration over the conjugate distance $x$ of the exchanged momentum $Q$ in the Fourier transform~\cite{DiVecchia:2022piu}. Hence, while there can exist finite frequency graviton contributions to the dressing, in \ref{s.eik}, we consider only contributions from low-frequency gravitons resulting from the soft factor for all eikonal amplitudes.

The leading order $\kappa$ coherent dressing results from an imaginary contribution to the eikonal phase at two loops. Given the $\kappa^{2}$ dependence in $\Delta$, we expect the above dressing to correspond to three-loop (and higher) eikonal amplitudes. However, inelastic eikonal amplitudes to these orders have thus far not been derived. We may nevertheless use \ref{s.eik} to derive radiative gravitational wave observables following the prescription for soft dressings in \cite{DiVecchia:2022owy, DiVecchia:2022nna}. In this section, we argue that classical radiative observables can be expressed in terms of the expectation values of the corresponding operators in a coherent state. In particular, this coherent state contains the $\mathcal{O}(\kappa^3)$ corrections in its exponential factor, which can be used to derive higher PM gravitational wave observables.

We begin by applying the BCH formula on \ref{dress} to factorize it into a product of exponential dressings involving single and double graviton modes \footnote{Formally, there is also a c-number exponential factor, i.e. a term with no gravitons. We derive it explicitly in Appendix \ref{A} and show that it vanishes.}.
\begin{align}
\exp \left[-\Delta\right] =  \exp\left[-\Delta_1 \right]\exp\left[-\Delta_2 \right] \,,
\label{dress.fact}
\end{align}
where the subscript $k$ of $\Delta_k$ denotes the number of dressing gravitons.  $\Delta_2$ is the 2-mode $\kappa^2$ operator in \ref{d1k3.f}.  The $\Delta_1$ term, on the other hand, is a 1-mode coherent state containing terms to all odd orders in $\kappa$, and its explicit form is provided in Appendix \ref{A}. The 1-mode coherent term $\Delta_1$ up to its $\kappa^3$ correction is 
\begin{align}
\Delta_1 &= \tilde{\Delta}_1 + \Delta_1^{\kappa^3} + \mathcal{O}(\kappa^5)\notag\\
\Delta_1^{\kappa^3} &= \frac{1}{2 \hbar} \int_{\vec{k}} d^3k \int_{\vec{l}} d^3l \left[a^{\dagger}_i (k) \left(A_{ij}(k,l) f^*_j(l) + B^*_{ij}(k,l) f_j(l)\right)    \right. \notag\\
& \left. \qquad \qquad  \qquad \qquad \qquad - a_i (k) \left(A^*_{ij}(k,l) f_j(l) + B_{ij}(k,l) f^*_j(l)\right) \right] \label{d1k3.f} \,,
\end{align}
with $\tilde{\Delta}_1$ as in \ref{d1.f} and $\Delta_1^{\kappa^3}$ being the $\kappa^3$ correction to the {\it coherent dressing}.

The factorized dressing \ref{dress.fact} allows us to identify relevant contributions for classical observables. The dressing is similar to squeezed coherent states \cite{GerryKinght}, generalized to continuous frequency modes and a 2-mode squeezing operator. One might thus expect a contribution from $\Delta_2$ for certain classical observables. However, for radiative observables considered in this paper, we establish that contributions from the 2-mode dressing $\Delta_2$ are $\mathcal{O}(\hbar)$ due to the normal ordering, and hence can be disregarded if only classical observables are concerned. Therefore, while \ref{dress} and \ref{dress.fact} are equivalent, the latter manifests the NE double graviton contributions in the gravitational dressing into a 1-mode coherent form relevant for classical observables to all odd powers in $\kappa$.

For an operator $\mathcal{Q}$ comprising only graviton modes, we have the identity
\begin{equation}
S^{\dagger} \mathcal{Q} S =  e^{\Delta} \mathcal{Q} e^{- \Delta} = \mathcal{Q} + \left[\Delta \,, \mathcal{Q}\right] + \frac{1}{2} \left[\Delta\,, \left[\Delta\,, \mathcal{Q}\right]\right] + \cdots \,.
\label{op.bch}
\end{equation}
Considering \ref{op.bch} with \ref{dress} on graviton creation and annihilation operators, we have
\begin{align}
e^{\Delta}a_i(k) e^{-\Delta} &=  a_i(k) + f_i(k) - \frac{1}{2}\int_{\vec{l}} d^3l \, \left(A_{ij}(k\,,l) f^*_{j} (l) + B^*_{ij}(k\,,l) f_{j} (l)\right) \notag\\
& \qquad- \int_{\vec{l}} d^3l \, \left(A_{ij}(k\,,l)a^{\dagger}_j(l) + B^*_{ij}(k\,,l) a_j(l)\right) + \mathcal{O}\left(\kappa^5\right) \label{ag2.ev1}\\
 e^{\Delta}a^{\dagger}_i(k) e^{-\Delta} &= a^{\dagger}_i(k) + f^*_i(k) - \frac{1}{2}\int_{\vec{l}} d^3l \, \left(A^*_{ij}(k\,,l) f_{j} (l) + B_{ij}(k\,,l) f^*_{j} (l)\right) \notag\\
& \qquad - \int_{\vec{l}} d^3l \, \left(A^*_{ij}(k\,,l)a_j(l) + B_{ij}(k\,,l) a^{\dagger}_j(l)\right) + \mathcal{O}\left(\kappa^5\right) \,, \label{adg2.ev1}
\end{align}
in which the terms involving combinations of $f_i$ with $A_{ij}$ and $B_{ij}$ result from the successive action of the single graviton mode over the double graviton mode operator from the unfactorized dressing \ref{dress}. 
This is equivalent to the result from the factorized dressing \ref{dress.fact} -- the first lines of \ref{ag2.ev1} and \ref{adg2.ev1} come from the 1-mode coherent dressing $\Delta_1$ in \ref{dress.fact}, while the second lines are from the two-mode dressing $\Delta_2$. 
A general expression of the above transformations on graviton creation and annihilation modes to all orders in $\kappa$ is provided in Appendix \ref{A}.

We denote the graviton vacuum as $\vert 0 \rangle$, and the dressed graviton vacuum by $e^{-\Delta} \vert 0 \rangle$. Classical gravitational wave observables following known approaches are associated with the expectation value of $\mathcal{Q}$ in the out-state $S \vert 0 \rangle$. Due to our consideration of graviton mode operator $\mathcal{Q}$ and the fact that $\text{Re} 2 \delta$ is a c-number phase, this reduces to its expectation value with respect to the dressed graviton vacuum, which we denote by $\langle \mathcal{Q} \rangle_{\Delta}$, i.e., 
\begin{align}
\langle 0 \vert S^{\dagger} \mathcal{Q} S \vert 0 \rangle = \langle 0 \vert e^{\Delta} \mathcal{Q} e^{-\Delta} \vert 0 \rangle := \langle \mathcal{Q} \rangle_{\Delta} \;. 
\label{op.ev}
\end{align}
We will also define expectation values with respect to the single graviton dressed state, i.e., coherent state, as
\begin{align}
\langle \mathcal{Q} \rangle_{\Delta_1} &:= \langle 0 \vert e^{\Delta_1} \mathcal{Q} e^{-\Delta_1} \vert 0 \rangle \;.
\label{op1.ev}
\end{align}

In the following, we establish that all classical observables associated with graviton mode operators $\mathcal{Q}$ satisfy $\langle \mathcal{Q} \rangle_{\Delta} = \langle \mathcal{Q} \rangle_{\Delta_1} + \mathcal{O}(\hbar)$. As we do not consider hard particle operators, the expectation values of $\mathcal{Q}$ will be taken to be the corresponding radiative observable. Let us now consider specific cases of radiative observables, such as the waveform represented by linear combinations of single graviton modes. For such observables, we can directly consider the expectation values of \ref{ag2.ev1} and \ref{adg2.ev1} to find
\begin{align}
\langle a_i(k)  \rangle_{\Delta} &= f_i(k) - \frac{1}{2}\int_{\vec{l}} d^3l \, \left(A_{ij}(k\,,l) f^*_{j} (l) + B^*_{ij}(k\,,l) f_{j} (l)\right) + \mathcal{O}\left(\kappa^5\right) = \langle a_i(k)  \rangle_{\Delta_1}\;, \label{ag2.ev}\\
\langle  a^{\dagger}_i(k) \rangle_{\Delta} &= f^*_i(k) - \frac{1}{2}\int_{\vec{l}} d^3l \, \left(A^*_{ij}(k\,,l) f_{j} (l) + B_{ij}(k\,,l) f^*_{j} (l)\right) + \mathcal{O}\left(\kappa^5\right) = \langle  a^{\dagger}_i(k) \rangle_{\Delta_1}\,, \label{adg2.ev}
\end{align}
which are entirely determined by the coherent contributions in the first lines of \ref{ag2.ev1} and \ref{adg2.ev1}, with the double graviton dressing corrections included, as expected. Later, these expectation values will be adopted to evaluate the waveforms of the emitted soft gravitons. 

We next consider the following operator involving two graviton modes,
\begin{equation}
\mathcal{Q}_2 = \int_{\vec{k}} d^3k \; C(k) a^{\dagger}_i(k) a_i(k) \,,  
\label{q2.op}
 \end{equation} 
with $C(k)$ a function of the wavevector $k$. The graviton number operator is realized with $C(k) = \hbar^{-1}$, while the radiated momentum for a single graviton results from $C(k) = k^{\mu}$ in our conventions.

The expectation value of \ref{q2.op} can now be determined using \ref{ag2.ev1} and \ref{adg2.ev1}. We find
\begin{align}
\langle \mathcal{Q}_2\rangle_{\Delta} & = \langle \mathcal{Q}_2\rangle_{\Delta_1} + \langle \mathcal{Q}_2\rangle_{\Delta}^{\text{Remainder}}\;, \label{q2.d}\\
\langle \mathcal{Q}_2\rangle_{\Delta_1} & = \int_{\vec{k}} d^3k \; C(k) \langle a^{\dagger}_i(k)\rangle_{\Delta_1} \langle a_i(k) \rangle_{\Delta_1}\;, \notag\\
&= \int_{\vec{k}} d^3k\; C(k) \left(f^*_i(k)f_i(k) - \frac{1}{2}\int_{\vec{l}} d^3l \, \left(f_i(k) A^*_{ij}(k\,,l) f_{j} (l) + f_i(k) B_{ij}(k\,,l) f^*_{j} (l) \right. \right. \notag\\
& \left. \left. \phantom{- \int_{\vec{l}} d^3l} \qquad \qquad    + f^*_i(k)A_{ij}(k\,,l) f^*_{j} (l) + f^*_i(k) B^*_{ij}(k\,,l) f_{j} (l)\right) \right) + \mathcal{O}\left(\kappa^5\right) \label{q2.d1} \;, \\
\langle \mathcal{Q}_2\rangle_{\Delta}^{\text{Remainder}} &=  \Bigg \langle \int_{\vec{k}} d^3k\; C(k)  \left(a^{\dagger}_i(k)a_i(k) - \int_{\vec{l}} d^3l \, \left(A^*_{ij}(k\,,l)a_j(l) a_i(k) + B_{ij}(k\,,l) a^{\dagger}_j(l) a_i(k) \right. \right. \notag \\
&\left. \left. \qquad \qquad \phantom{- \int_{\vec{l}} d^3l} + A_{ij}(k\,,l)a^{\dagger}_i(k) a^{\dagger}_j(l) + B^*_{ij}(k\,,l) a^{\dagger}_i(k) a_j(l)\right) \right) \Bigg \rangle + \mathcal{O}\left(\kappa^5\right)  \,. \label{q2.d2} 
\end{align}
The (quantum) remainder $\langle \mathcal{Q}_2\rangle_{\Delta}^{\text{Remainder}}$ of \ref{q2.d2} is non-vanishing, however, it is sub-leading in $\mathcal{O} (\hbar)$ when compared to $\langle \mathcal{Q}_2\rangle_{\Delta_1}$ of \ref{q2.d1}. The additional $\hbar$ factor arises from normal ordering when evaluating \ref{q2.d2} through the commutation relation for the mode operators in \ref{mode.comm} and \ref{del.def}. Hence, in the classical limit with $\hbar \rightarrow 0$, only $\langle \mathcal{Q}_2\rangle_{\Delta_1}$ contributes, which is from the 1-mode dressing $\Delta_1$. For the examples mentioned previously, when $\mathcal{Q}_2$ is the graviton number operator, $\langle \mathcal{Q}_2\rangle_{\Delta_1} \sim \hbar^{-1}$ diverges in the classical limit as expected for coherent graviton emissions. For $\mathcal{Q}_2$ with $C(k)=k^{\mu}$, $\langle \mathcal{Q}_2\rangle_{\Delta_1} \sim \hbar^0$ gives the radiated momentum of the soft gravitons. In both cases, we see that the double graviton dressing functions $A_{ij}, B_{ij}$ and their complex conjugates provide $\kappa^4$ corrections through their contractions with the leading Weinberg soft factor $f_{i}$.

We lastly address the expectation value of \ref{s.eik} with respect to the graviton vacuum,
\be
\langle 0 \vert S \vert 0 \rangle =  \langle 0 \vert e^{-\Delta_1} e^{-\Delta_2} \vert 0 \rangle e^{i \text{Re} 2 \delta}:= e^{- \text{Im} 2 \delta}e^{i \text{Re} 2 \delta}\;.
\ee
The expectation value of the dressing S-matrix is non-vanishing and can be interpreted as the imaginary contribution to the eikonal phase, i.e., $\langle 0 \vert S \vert 0 \rangle = e^{i 2 \delta}$ with $2 \delta = \text{Re} 2 \delta + i \text{Im} 2 \delta$~\cite{DiVecchia:2022nna}.  The result for $\text{Im} 2 \delta$  follows from the normal ordering of the operators in $\Delta_1$ and $\Delta_2$, and along with the BCH formula  we find
\begin{align}
\text{Im} 2 \delta &= \text{Im} 2 \delta_{\kappa^2} + \text{Im} 2 \delta_{\kappa^4} + \mathcal{O}(\kappa^5) + \mathcal{O}(\hbar^0)\;, \label{im2d}\\
\text{Im} 2 \delta_{\kappa^2} &= \frac{1}{2 \hbar}\int_{\vec{k}} d^3k f_i^*(k)f_i(k) \;,\label{im2d.k1}\\
\text{Im} 2 \delta_{\kappa^4} &= - \frac{1}{4 \hbar} \int_{\vec{k}} d^3k \int_{\vec{l}} d^3l \left(f_i(k) A^*_{ij}(k,l) f_j(l) + f_i(k) B_{ij}(k,l) f^*_j(l)\right.   \notag\\
& \left. \phantom {-\frac{1}{2 \hbar}\int_{\vec{k}} d^3k } \quad + f^*_i(k) A_{ij}(k,l) f^*_j(l) + f^*_i(k)B^*_{ij}(k,l) f_j(l)\right)\;, \label{im2d.k3}
\end{align}
with $\text{Im} 2 \delta_{\kappa^2}$ in \ref{im2d.k1} the $\kappa^2$ contribution from two Weinberg soft gravitons~\cite{DiVecchia:2022nna}, while $\text{Im} 2 \delta_{\kappa^4}$ of \ref{im2d.k3} is a $\kappa^4$ contribution from the coherent dressing due to the contraction of the double graviton with two Weinberg soft gravitons. In \ref{im2d}, the $\mathcal{O}(\kappa^5)$ terms are subleading $\kappa$ contributions from the coherent dressing, while $\mathcal{O}(\hbar^0)$ terms are the superclassical contributions from $e^{-\Delta_2}$. The $\mathcal{O}(\hbar^0)$ contributions can likely be identified with super-Poissonian statistics. On the other hand, the $\hbar^{-1}$ terms in \ref{im2d} contribute to the usual Poissonian statistics associated with a coherent dressing along the lines discussed in \cite{DiVecchia:2022nna,Britto:2021pud}
\begin{align}
& \mathcal{P}_N = \frac{1}{N! \hbar^N} \Big\langle \prod_{i=1}^N \int_{\vec{k}} d^3k_i a_i^{\dagger} a_i \Big\rangle_{\Delta} = \frac{1}{N!} \left[2 \text{Im} 2 \delta\right]^N e^{-2 \text{Im} 2 \delta}\;, \notag\\
& \sum_{N=0}^{\infty} N \mathcal{P}_N  = 2 \text{Im} 2 \delta = \sum_{N=0}^{\infty} N^2 \mathcal{P}_N - \left(\sum_{N=0}^{\infty} N \mathcal{P}_N\right)^2\;.
\label{poiss}
\end{align}
Due to the presence of $\kappa^4$ (and $\mathcal{O}(\kappa^5)$) corrections in $\text{Im} 2 \delta$, we now expect a relationship with $\mathcal{O}(G^2)$ emitted momentum and energy spectrum. We will return to this in the following section.

\section{Gravitational wave observables}

We now address and evaluate specific gravitational wave observables resulting from \ref{dress.fact} for scattering events of scalar particles with a soft graviton dressing. For the external massive particles, we adopt an all-outgoing convention with the following parametrization
\begin{align}
p^{\mu}_a &= \eta_a \left(E_a\,, \vec{p}_a \right) \,,\label{p.mom}
\end{align}
with $\eta_a = +1$ ($\eta_a = -1$) for outgoing (ingoing) massive particles and $p_a^2 = -m_a^2$.  The graviton momentum will be parametrized as $k^{\mu} = \omega_k q_{k}^{\mu}$ as given in \ref{mom.param}. This results in
\begin{align}
p_a\cdot k &= \eta_a \omega_k (\eta_a p_a\cdot q_k) \,,\notag\\
\eta_a p_a\cdot q_k &= - (E_a - \vec{p}_a\cdot\hat{k})\,.
\label{pk.def}
\end{align} 

For two external particles, we can construct the following Lorentz invariant quantities 
\begin{equation}
\sigma_{ab} = - \eta_a \eta_b \frac{p_a\cdot p_b}{m_a m_b}\,, \qquad \Delta_{ab}={\cosh^{-1}\sigma_{ab} \over \sqrt{\sigma^2_{ab}-1}}\;,
\label{rv.def}
\end{equation}
and our results will be expressed in terms of them. In evaluating the classical gravitational-wave observables from expectation values, one will also deal with contractions over graviton polarization tensors that form the transverse and traceless projection operator defined by 
\begin{align}
\Pi^{\mu \nu \alpha \beta} (\hat{k}) &= \varepsilon_{i}^{\mu \nu}(k) \varepsilon_{i}^{* \alpha \beta}(k) =  \varepsilon_{i}^{* \mu \nu}(k) \varepsilon_{i}^{\alpha \beta}(k) \;,\label{proj.def}\\
&=  \frac{1}{2}\left[\Pi^{\mu \alpha}(\hat{k}) \Pi^{\nu \beta}(\hat{k}) + \Pi^{\nu \alpha}(\hat{k}) \Pi^{\mu \beta}(\hat{k}) -  \Pi^{ \mu \nu}(\hat{k})\Pi^{\alpha \beta}(\hat{k}) \right]\;,  \label{proj.d}\\
\Pi^{\mu \nu}(\hat{k}) &= \eta^{\mu \nu} + \lambda^{\mu} k^{\nu} + \lambda^{\nu} k^{\mu}  = \Pi^{\nu \mu} (\hat{k}) \;, \label{proj.p}\\
& \qquad  \lambda^2 = 0 = k^2\;; \quad \lambda\cdot k = -1\;, \label{proj.onshell}
\end{align}
with $\lambda^{\mu}$ a reference vector. This projection operator, by definition, depends on the orientation $\hat{k}$ of the on-shell gravitons, which obeys \ref{proj.onshell}. 
Moreover, it is transverse to the momentum and reference vector, i.e., 
\begin{equation}
k^{\mu}\Pi_{\mu \nu} (\hat{k}) = 0 = \lambda^{\mu}\Pi_{\mu \nu} (\hat{k})\,.
\label{pr.tr}
\end{equation}
Due to the collinear limit in the double graviton dressing, all projection operators in the final result are along a single orientation, which we take to be $\hat{k}$.  Thus, we can express $F^{\mu \nu}$ in \ref{f.def}, and $A^{\mu \nu \rho \sigma}$ and $B^{\mu \nu \rho \sigma}$ in \ref{nAB.def} as dependent on the graviton orientation $\hat{k}$ and the energies $\omega_k$ and $\omega_l$. To evaluate observables constructed out of these quantities, we integrate over these kinematic variables of the dressing gravitons to find results that only depend on the external particle momenta \ref{p.mom}, their relativistically invariant combinations \ref{rv.def} and the total graviton cut-off frequency $\omega$.

Some radiative observables, such as the waveform and angular momentum, 
are sensitive to the integration around the poles of $\omega_{k,l}$. To fix the related causality issue as suggested in~\cite{DiVecchia:2022owy}, we adopt the Feynman $i0$ prescription by adding $-i0_{k,l}$ ($+i0_{k,l}$) to the $\omega_{k,l}$ appearing with the polarization $\varepsilon^*$ ($\varepsilon$). From the definitions of $F^{\mu\nu}$, $A^{\mu\nu\alpha\beta}$ and $B^{\mu\nu\alpha\beta}$ as given in \ref{f.def} and \ref{nAB.def}, we define the following associated polarization-projected soft dressing quantities $f_{\mu\nu}$, $\alpha_{\mu\nu}^{\phantom {\mu \nu} \alpha  \beta}$ and $\beta_{\mu\nu}^{\phantom {\mu \nu} \alpha  \beta}$, respectively, 
\begin{align}
& f_{\mu \nu} (\omega_k, \hat{k}) :=  \Pi_{\mu \nu \alpha  \beta}(\hat{k}) F^{\alpha \beta}(\omega_k, \hat{k}) = \kappa \Pi_{\mu \nu \alpha  \beta}(\hat{k})\sum_{n} \frac{1}{\eta_n \omega_k + i0_k}  \frac{p_n^{\alpha} p_n^{\beta}}{\eta_n p_n\cdot q_k}\;, \label{f1.def}\\
& (2 \pi)^2 \delta(\Omega_{k}\,, \Omega_{l}) \alpha_{\mu\nu}^{\phantom {\mu \nu} \alpha  \beta}(\omega_k\,, \omega_l \,, \hat{k}):=\Pi_{\mu \nu \rho  \sigma}(\hat{k}) A^{\rho \sigma \alpha \beta}(\omega_k, \omega_l, \hat{k}) \notag\\
&\qquad \qquad  = (2 \pi)^2\delta(\Omega_{k}\,, \Omega_{l}) \kappa^2 \Pi_{\mu \nu \rho  \sigma}(\hat{k})    \sum_{n} \frac{(p_n^{\rho} p_n^{\alpha} \eta^{\beta \sigma} + p_n^{\rho} p_n^{\beta} \eta^{\alpha \sigma} + p_n^{\sigma} p_n^{\alpha} \eta^{\beta \rho} + p_n^{\sigma} p_n^{\beta} \eta^{\alpha \rho})}{(\eta_n (\omega_k + \omega_l) + i0_k +i0_l) (\eta_n p_n\cdot q_k) } \;, \label{al.def}\\
& (2 \pi)^2 \delta(\Omega_{k}\,, \Omega_{l}) \beta_{\mu\nu}^{\phantom {\mu \nu} \alpha  \beta}(\omega_k\,, \omega_l \,, \hat{k})=\Pi_{\mu \nu \rho  \sigma}(\hat{k}) B^{\rho \sigma \alpha \beta}(\omega_k, \omega_l, \hat{k})   \notag\\
&\qquad \qquad =  (2 \pi)^2 \delta(\Omega_{k}\,, \Omega_{l}) \kappa^2 \Pi_{\mu \nu \rho  \sigma}(\hat{k})  \sum_{n} \frac{(p_n^{\rho} p_n^{\alpha} \eta^{\beta \sigma} + p_n^{\rho} p_n^{\beta} \eta^{\alpha \sigma} + p_n^{\sigma} p_n^{\alpha} \eta^{\beta \rho} + p_n^{\sigma} p_n^{\beta} \eta^{\alpha \rho})}{(\eta_n (\omega_k - \omega_l) - i0_k - i0_l)(\eta_n p_n\cdot q_k)} \;.\label{bet.def}
\end{align}
The collinear $\delta(\Omega_{k}\,, \Omega_{l})$ appearing above reduces the double angular integrals to the one just over $\Omega_k$ when evaluating the observables. Moreover, these quantities satisfy the following relations, 
\begin{align}
f^*_{\mu \nu} (-\omega_k, \hat{k}) &= -f_{\mu \nu} (\omega_k, \hat{k}) \;,\notag\\
\alpha^*_{\mu\nu \alpha  \beta}(-\omega_k\,, -\omega_l \,, \hat{k}) &= -\alpha_{\mu\nu \alpha  \beta}(\omega_k\,, \omega_l \,, \hat{k}) \;,\notag\\
\beta^*_{\mu\nu \alpha  \beta}(-\omega_k\,, -\omega_l \,, \hat{k}) &= -\beta_{\mu\nu \alpha  \beta}(\omega_k\,, \omega_l \,, \hat{k})\;.
\label{fab.id}
\end{align}

In addition, we define the following quantities, denoted by $l^{\pm}_{\mu\nu}$, which appear due to $\Delta_1^{\kappa^3}$ and thus on many occasions in later discussions, 
\begin{align}
 l^+_{\mu \nu}(\omega_k\,, \omega_l \,, \hat{k})& :=\alpha_{\mu\nu \rho \sigma}(\omega_k\,, \omega_l \,, \hat{k}) f^{*\,\rho \sigma} (\omega_l \,, \hat{k}) = \Pi_{\mu \nu \rho \sigma} (\hat{k}) C^{+ \rho \sigma}(k,l) \;,\notag\\
l^-_{\mu \nu}(\omega_k\,, \omega_l \,, \hat{k}) & :=\beta^*_{\mu\nu \rho \sigma}(\omega_k\,, \omega_l \,, \hat{k}) f^{\rho \sigma} (\omega_l \,, \hat{k})   =  \Pi_{\mu \nu \rho \sigma} (\hat{k}) C^{- \rho \sigma}(k,l)\;,
\label{lpm}
\end{align}
where\footnote{Note that \ref{Cpm.def} has the following transformation on changing the sign of $\omega_l$ 
\begin{equation}
C^{\pm \alpha \beta}(k\,,l) \xrightarrow{\omega_l \to -\omega_l} -C^{\mp \alpha \beta}(k\,,l)\;. \nn
\label{cpm.l}
\end{equation}}
\begin{align}
C^{\pm \, \alpha \beta}(k,l) &:= \kappa^3 \sum_{n,i} N^{\alpha \beta}_{ni}(\hat{k}) \frac{1}{\left(\eta_n (\omega_k \pm \omega_l) + i0_k + i0_l\right) \left(\eta_i \omega_l  \mp i0_l\right)}\;, \label{Cpm.def} \\
N^{\alpha \beta}_{ni}(\hat{k})& := \frac{(p_n^{\rho} p_n^{\alpha} \eta^{\beta \sigma} + p_n^{\rho} p_n^{\beta} \eta^{\alpha \sigma} + p_n^{\sigma} p_n^{\alpha} \eta^{\beta \rho} + p_n^{\sigma} p_n^{\beta} \eta^{\alpha \rho})}{\eta_n p_n\cdot q_k} P_{\rho \sigma \gamma \delta}\frac{p_i^{\gamma} p_i^{\delta}}{\eta_i p_i\cdot q_k}\;, 
\label{cpm}
\end{align}
and with 
\begin{equation}
P_{\gamma \delta \rho \sigma} = \frac{1}{2}\left(\eta_{\gamma \rho} \eta_{\delta \sigma} + \eta_{\gamma \sigma} \eta_{\delta \rho} - \eta_{\gamma \delta} \eta_{\rho \sigma}\right)
\label{Pp.def}
\end{equation}
as the $\lambda^{\mu}$-independent part of the projection operator. Hence, the terms appearing in the sum of \ref{Cpm.def} are effectively fixed with respect to the de Donder gauge. 

We highlight a notational choice introduced in \ref{Cpm.def} which we will implement for the remainder of the paper. The label $n$ will denote the external particle with the double graviton vertex, while labels $i$ and $j$ will be reserved for particles on which Weinberg soft gravitons are attached. In the following, we carry out the sum over all external particles with each label considered generally distinct. However, the collinear condition on the double graviton vertex will restrict the sum to specific limits of the kinematic variables of the external particles they attach. We will return to this limiting procedure in the discussion section of this paper. In the following subsections, we derive general results for the waveform, emitted momentum spectrum and angular momentum.

\subsection{Waveform of soft bremsstrahlung and memory effect} \label{mem}

From the metric in \ref{met}, the waveform of soft gravitons is given by the expectation value $W_{\mu \nu}(x) := 2 \kappa \langle h_{\mu \nu}(x)\rangle_{\Delta}$, with the graviton mode given by \ref{gm}
\begin{equation}
h_{\mu \nu} (x) = \int_{\vec{k}} d^3k \left[ a_i (k) \varepsilon_{i\,,\mu \nu} (k) e^{ik\cdot x} + a^{\dagger}_i (k) \varepsilon^*_{i\,,\mu \nu} (k) e^{-ik\cdot x} \right]\;. \label{gm.2}
\end{equation}

Being interested in the waveform observed at asymptotic infinity, we consider the spacetime coordinates as $x = \left( u+r \,, r \hat{x}\right)$, with $u$ the retarded time, $r$ the radial distance and $\hat{x}$ the angular orientation of the detector placed at large $r$. Using \ref{ag2.ev} and \ref{adg2.ev} in \ref{gm.2}, and subsequently using the definitions given in \ref{f1.def}, \ref{al.def} and \ref{bet.def}, we find
\begin{align}
W_{\mu \nu}(x)  &= W^G_{\mu \nu}(x) + W^{G^2}_{\mu \nu}(x)\;, \notag\\
W^G_{\mu \nu}(x) &= \frac{\kappa}{(2 \pi)^3} \int_{0}^{\infty} d \omega_k \omega_k \oint  d\Omega_k  \,\Theta\left(\omega - \omega_k \right) \notag\\
&\qquad \qquad \qquad \qquad  \left(e^{-i\omega_k u - i \omega_k r(1 - \hat{k}\cdot \hat{x})} f_{\mu \nu}(\omega_k\,, \hat{k})  +  e^{i\omega_k u + i \omega_k r(1 - \hat{k}\cdot \hat{x})} f^*_{\mu \nu}(\omega_k\,, \hat{k}) \right)\;, \label{mem.int}\\
W^{G^2}_{\mu \nu}(x) & = - \frac{\kappa}{4 (2 \pi)^4} \int_{-\infty}^{\infty} d \omega_k \omega_k \Theta(\omega_k)   \int_{-\infty}^{\infty} d \omega_l \omega_l  \Theta(\omega_l) \oint  d\Omega_k  \,   \Theta\left(\omega - \omega_k -  \omega_l \right) \notag\\
& \qquad \qquad \qquad  \left[ e^{-i\omega_k u - i \omega_k r(1 - \hat{k}\cdot \hat{x})} \left( l^+_{\mu\nu}(\omega_k\,, \omega_l \,, \hat{k}) + l^{-}_{\mu\nu}(\omega_k\,, \omega_l \,, \hat{k})\right) \phantom{ - \kappa \int_{\vec{k}} d^3k \int_{\vec{l}} d^3l}\right. \notag\\
&\left. \phantom{\int_{\vec{k}} }\qquad \qquad \qquad \qquad  + e^{i\omega_k u + i \omega_k r(1 - \hat{k}\cdot\hat{x})} \left( l^{+*}_{\mu\nu}(\omega_k\,, \omega_l \,, \hat{k}) + l^{-*}_{\mu\nu}(\omega_k\,, \omega_l \,, \hat{k}) \right) \right]\,.
\label{mem.int2}
\end{align}
The Heaviside step functions in \ref{mem.int} and \ref{mem.int2} restrict the $\omega_{k,l}$ integrals to the soft region \footnote{For the two graviton emission case, we have the step function $\Theta(\omega-\omega_k-\omega_l)$ to restrict the integration bounded by the soft region. In later manipulations, we further use $\Theta(\omega_k)$ to rewrite the $\omega_k$ integral as
\begin{equation}
\int_{0}^{\infty} d \omega_k = \int_{-\infty}^{\infty} d \omega_k \Theta(\omega_k) \,,
\end{equation}
and likewise for the $\omega_l$ integral. 
}. The explicit use of the step function is not necessary in the case of \ref{mem.int}, since the integral picks up the contribution from $\omega_k \to 0$ and was evaluated in~\cite{DiVecchia:2022nna,DiVecchia:2022owy}. This provides the leading contribution to the linear memory effect. The evaluation of \ref{mem.int2} will provide the next-to-eikonal correction of \ref{mem.int} to ${\cal O}(G^2)$ and its corresponding memory effect. We note that the indicated counting in powers of $G$ follows conventions for PM waveforms from amplitudes. More specifically, \ref{mem.int} and \ref{mem.int2} would respectively correspond to soft contributions to the post-linear (order $G^2$) and post-post-linear contribution (order $G^3$) terms of the classical PM waveform \footnote{The term at order $G$ in the classical PM waveform is time independent and does not contribute to the memory effect~\cite{Damour:2020tta}. We refer the reader to~\cite{Bini:2023fiz} for further comparisons between classical PM waveforms and waveforms from amplitudes. }

As the soft graviton dressing is an IR effect, we will measure the corresponding waveform in the large $r$ limit, for which the angular integration can be carried out by saddle point approximation, e.g., see \cite{DiVecchia:2022owy}. Denoting the angular position of the detector relative to the source by $\hat{x}$, the waveform \ref{mem.int2} at $\mathcal{O}(G^2)$ can be reduced to 
\begin{align}
W^{G^2}_{\mu \nu}(x) & = - \frac{\kappa}{(2 \pi)^2} \frac{1}{4  r} \int_{-\infty}^{\infty} \frac{d \omega_k}{2 \pi i} \Theta(\omega_k)   \int_{-\infty}^{\infty} d \omega_l \omega_l  \Theta(\omega_l) \Theta\left(\omega - \omega_k - \omega_l \right) \notag\\
& \left[ e^{-i\omega_k u} \left( l^+_{\mu\nu}(\omega_k\,, \omega_l \,, \hat{x}) + l^{-}_{\mu\nu}(\omega_k\,, \omega_l \,, \hat{x}) \right)  - e^{i\omega_k u} \left( l^{+*}_{\mu\nu}(\omega_k\,, \omega_l \,, \hat{x}) + l^{-*}_{\mu\nu}(\omega_k\,, \omega_l \,, \hat{x})\right) \right]
\label{mem.int4}
\end{align}
with the graviton now oriented along the detector $k^{\mu} = \omega_k (1\,,\hat{x})$, as required by the angular saddle. From the transformations noted in \ref{fab.id} and the definitions in \ref{lpm}, we have
\begin{align}
l^{+*}_{\mu\nu}(-\omega_k\,, -\omega_l \,, \hat{x}) + l^{-*}_{\mu\nu}(-\omega_k\,, -\omega_l \,, \hat{x})= l^+_{\mu\nu}(\omega_k\,, \omega_l \,, \hat{x}) + l^-_{\mu\nu}(\omega_k\,, \omega_l \,, \hat{x})\,.
\label{trans}
\end{align}
Hence, considering $\omega_k \to -\omega_k$ and $\omega_l \to - \omega_l$ in the last term of the last line of \ref{mem.int4}, we get
\begin{align}
W^{G^2}_{\mu \nu}(x) & = - \frac{\kappa}{(2 \pi)^2} \frac{1}{4 r} \int_{-\infty}^{\infty} \frac{d \omega_k}{2 \pi i}    \int_{-\infty}^{\infty} d \omega_l \omega_l \,   \Theta\left(\vert\omega\vert - \vert\omega_k\vert -  \vert\omega_l \vert\right) e^{-i\omega_k u}  \notag\\
& \qquad \qquad \left(l^+_{\mu\nu}(\omega_k\,, \omega_l \,, \hat{x}) + l^-_{\mu\nu}(\omega_k\,, \omega_l \,, \hat{x})\right) \left(\Theta(\omega_k) \Theta(\omega_l) + \Theta(-\omega_k) \Theta(-\omega_l)\right)
\label{mem.int5}
\end{align}

From the expressions for $l^{\pm}_{\mu \nu}$ in \ref{lpm} and using \ref{pk.def} we find
\begin{align}
&l^+_{\mu\nu}(\omega_k\,, \omega_l \,, \hat{x}) + l^-_{\mu\nu}(\omega_k\,, \omega_l \,, \hat{x}) \notag\\
& =\Pi_{\mu \nu \alpha \beta}(\hat{x}) \kappa^3 \sum_{n,i} \frac{(p_n^{\gamma} p_n^{\alpha} \eta^{\beta \delta} + p_n^{\gamma} p_n^{\beta} \eta^{\alpha \delta} + p_n^{\delta} p_n^{\alpha} \eta^{\beta \gamma} + p_n^{\delta} p_n^{\beta} \eta^{\alpha \gamma})}{(E_n - \vec{p}_n\cdot \hat{x})} P_{\gamma \delta \rho \sigma}\frac{p_i^{\rho} p_i^{\sigma}}{(E_i - \vec{p}_i\cdot \hat{x})}\notag\\
&\left[ \frac{1}{\left(-\eta_n (\omega_k -\omega_l) - i0_k - i0_l\right) \left(-\eta_i \omega_l  - i0_l\right)} + \frac{1}{\left(-\eta_n (\omega_k + \omega_l) - i0_k - i0_l\right) \left(-\eta_i \omega_l  + i0_l\right)}\right]\,,
\label{la.int}
\end{align}
which follows from using the transversality property of the Weinberg soft factor and the projection operator in \ref{F.tr} and \ref{pr.tr}, respectively.
We can further simplify the $\omega_k$ and $\omega_l$ independent pieces in \ref{la.int} \footnote{Using \ref{Pp.def} and \ref{rv.def}, we have
\begin{align}
&(p_n^{\gamma} p_n^{\alpha} \eta^{\beta \delta} + p_n^{\gamma} p_n^{\beta} \eta^{\alpha \delta} + p_n^{\delta} p_n^{\alpha} \eta^{\beta \gamma} + p_n^{\delta} p_n^{\beta} \eta^{\alpha \gamma})P_{\gamma \delta \rho \sigma}p_i^{\rho} p_i^{\sigma} = - 2 \left(\eta_n \eta_i m_n m_i \sigma_{ni} \left(p_n^{\alpha} p_i^{\beta} + p_i^{\alpha} p_n^{\beta}\right) - m_i^2 p_n^{\alpha}p_n^{\beta}\right)\,.
\label{nump}
\end{align} 
}, so that the expression of the integrand of $W^{G^2}_{\mu \nu}(x)$ can be further reduced to
\begin{align}
&W^{G^2}_{\mu \nu}(x)  = \frac{8 G^2 }{r} \Pi_{\mu \nu \alpha \beta}(\hat{x}) \sum_{n,i} \frac{\left(\eta_n \eta_i m_n m_i \sigma_{ni} \left(p_n^{\alpha} p_i^{\beta} + p_i^{\alpha} p_n^{\beta}\right) - m_i^2 p_n^{\alpha}p_n^{\beta}\right)}{(E_n - \vec{p}_n\cdot\hat{x}) (E_i - \vec{p}_i\cdot \hat{x})} \notag\\
&\qquad  \int_{-\infty}^{\infty} \frac{d \omega_k}{2 \pi i}    \int_{-\infty}^{\infty} d \omega_l \omega_l   \,   \Theta\left(\vert\omega\vert - \vert\omega_k\vert -  \vert\omega_l \vert\right) \left(\Theta(\omega_k) \Theta(\omega_l) + \Theta(-\omega_k) \Theta(-\omega_l)\right) e^{-i\omega_k u}  \notag\\
& \left[ \frac{1}{\left(-\eta_n (\omega_k -\omega_l) - i0_k - i0_l\right) \left(-\eta_i \omega_l  - i0_l\right)} + \frac{1}{\left(-\eta_n (\omega_k + \omega_l) - i0_k - i0_l\right) \left(-\eta_i \omega_l  + i0_l\right)}\right]\;.
\label{mem.int6}
\end{align}

We will now perform the $\omega_k$ and $\omega_l$ integrals in the last two lines of \ref{mem.int6}. The $\omega_k$ integral can be evaluated by first rescaling $-\eta_n \omega_k \to \omega_k$ so that the integrand has simple poles in the upper half-plane. The residue theorem applied to the resulting $\omega_k$ integral then provides 
\begin{align}
&\int_{-\infty}^{\infty} \frac{d \omega_k}{2 \pi i}    \int_{-\infty}^{\infty} d \omega_l \omega_l   \,   \Theta\left(\vert\omega\vert - \vert \eta_n \omega_k\vert -  \vert\omega_l \vert\right) \left(\Theta(-\eta_n \omega_k) \Theta(\omega_l) + \Theta(\eta_n \omega_k) \Theta(-\omega_l)\right) e^{i \eta_n \omega_k u}  \notag\\
& \qquad \left[ \frac{1}{\left(\omega_k + \eta_n \omega_l - i0_k - i0_l\right) \left(-\eta_i \omega_l  - i0_l\right)} + \frac{1}{\left(\omega_k  - \eta_n \omega_l - i0_k - i0_l\right) \left(-\eta_i \omega_l  + i0_l\right)}\right] \;, \notag\\
\notag\\
&\qquad = \int_{-\infty}^{\infty} d \omega_l \omega_l   \,   \Theta\left(\vert\omega\vert - 2 \vert\omega_l \vert\right) \left[ (\Theta(\omega_l) \Theta(\omega_l) + \Theta(-\omega_l) \Theta(-\omega_l))\frac{e^{-i\omega_l u - \eta_n u 0_k}}{ \left(-\eta_i \omega_l  - i0_l\right)} \right. \notag\\
&\left. \qquad \qquad \qquad \qquad\qquad \qquad\qquad \qquad  + (\Theta(\omega_l) \Theta(-\omega_l) + \Theta(-\omega_l) \Theta(\omega_l))\frac{e^{i\omega_l u - \eta_n u 0_k}}{ \left(-\eta_i \omega_l  + i0_l\right)}\right]\;, \notag\\
&\qquad\qquad = \Theta(\eta_n u) \int_{-\frac{\omega}{2}}^{\frac{\omega}{2}} d \omega_l \omega_l  \frac{e^{-i\omega_l u}}{ \left(-\eta_i \omega_l  - i0_l\right)}\,,\label{k.int}
\end{align}
where in the last equality of \ref{k.int} we used the property that $\Theta(\omega_l) \Theta(-\omega_l) =0$, and used the step functions to express the $\omega_l$ integral symmetrically. The overall factor of $\Theta(\eta_n u)$ comes the requirement that $\eta_n u$ be positive in $e^{- \eta_n u 0_k}$. 
The $\omega_l$ integral in the last equality of \ref{k.int} can be evaluated using the identity
\begin{equation}
\frac{1}{\omega_l + c \pm i0} = \text{PV}\frac{1}{\omega_l + c} \mp i \pi \delta(\omega_l + c)\,,
\label{pvd}
\end{equation}
where PV denotes the principal part of the integral and $c$ a constant. We thus have 
\begin{equation}
\int_{-\frac{\omega}{2}}^{\frac{\omega}{2}} d \omega_l \omega_l  \frac{e^{-i\omega_l u}}{ \left(-\eta_i \omega_l  - i0_l\right)} = -\frac{2}{u} \eta_i  \sin \left(\frac{\omega u}{2}\right)\;. 
\label{l.int} 
\end{equation}

Hence, using \ref{k.int} and \ref{l.int} in \ref{mem.int6} gives us our final result
\begin{align}
W^{G^2; \alpha \beta}_{\text{TT}}(x) :=\Pi^{\mu \nu \alpha \beta}
 W^{G^2}_{\mu \nu}(\hat{x})  = - \frac{16 G^2 }{r u} \sin \left(\frac{\omega u}{2}\right) \sum_{n,i}  W_{n,i}^{\alpha\beta}(p_n, p_i;\hat{x}) \Theta(\eta_n u) \;,
\label{mem.int7}
\end{align}
with $W^{G^2; \alpha \beta}_{\text{TT}}(x)$ the transverse-traceless part of $W^{G^2}_{\mu \nu}(x)$, and $W_{n,i}^{\alpha\beta}(p_n, p_i;\hat{x})$ a relativistically invariant combination of the external momenta and orientation from the source to the detector
\begin{align}
 W_{n,i}^{\alpha\beta}(p_n, p_i;\hat{x}) &:= \frac{\eta_n m_n m_i \sigma_{ni} \left(p_n^{\alpha} p_i^{\beta} + p_i^{\alpha} p_n^{\beta}\right) -  \eta_i m_i^2 p_n^{\alpha}p_n^{\beta} }{(E_n - \vec{p}_n\cdot \hat{x}) (E_i - \vec{p}_i\cdot \hat{x})} \;.
\end{align}

It is instructive to consider \ref{mem.int7} in the small $\omega u$ limit to find
\begin{align}
W^{G^2; \alpha \beta}_{\text{TT}}(x)  &= - \frac{8 G^2 \omega }{r} \sum_{n,i}  W_{n,i}^{\alpha\beta}(p_n, p_i;\hat{x}) \Theta(\eta_n u) \left(1 + \sum_{l=1}^{\infty} {(-1)^l \over (2 l+1)!} \left({\omega u \over 2}\right)^{2l} \right)\,.
\label{mem.int8}
\end{align}
The second term in the parenthesis can be left small for all $u$ by an asymptotic double scaling limit: $\omega\rightarrow 0$ and $u\rightarrow \pm \infty$ but keeping $\omega u:= \pm \phi_0$ finite and small. A frame-independent definition of the memory effect follows from considering the difference of \ref{mem.int8} between the asymptotic future and asymptotic past~\cite{Veneziano:2022zwh}
\be
\Delta W^{G^2; \alpha \beta}_{\text{TT}}(x) := W^{G^2; \alpha \beta}_{\text{TT}}(x) \vert_{u \to \infty} - W^{G^2; \alpha \beta}_{\text{TT}}(x)\vert_{u \to -\infty}\;.
\label{mem.def}
\ee

We consider the $\mathcal{O}(G^2)$ gravitational memory effect \ref{mem.def} in the simplified case of $n = i$, where the kinematic factor simplifies to
\be  
W_{n,i}^{\alpha\beta}(p_n, p_i;\hat{x})  \xrightarrow{n=i} \frac{\eta_n m_n^2 p_n^{\alpha}p_n^{\beta}}{(E_n - \vec{p}_n\cdot \hat{x})^2}\;.
\ee
and hence apart from the $\mathcal{O}(\phi_0^{k\ge 2})$ corrections, the leading contribution to the $\mathcal{O}(G^2)$ gravitational memory effect from \ref{mem.int8} is 
\be
\Delta W^{G^2; \alpha \beta}_{\text{TT}}(x) = - \frac{8 G^2 \omega }{r} \sum_{n} \frac{m_n^2 p_n^{\alpha}p_n^{\beta}}{(E_n - \vec{p}_n\cdot \hat{x})^2}\;.
\label{mem.lead}
\ee

Before proceeding, we will briefly discuss the difficulty in comparing \ref{mem.int8} and \ref{mem.lead} with the current literature. Both results have an overall dependence on $\omega$ (the cut-off on the sum over graviton energies) as a consequence of integrating over the two frequencies in the double graviton contribution of the dressing. At present, the PM waveform is known up to one-loop~~\cite{Elkhidir:2023dco,Brandhuber:2023hhy,Herderschee:2023fxh,Georgoudis:2023lgf, Bini:2023fiz, Georgoudis:2023eke,Georgoudis:2023ozp,Bohnenblust:2023qmy,Bini:2024rsy}, which provides the leading correction to the previously known tree-level result~\cite{Jakobsen:2021smu,Mougiakakos:2021ckm,DeAngelis:2023lvf,Kovacs:1977uw}. We recall that the tree level and one-loop waveform results respectively correspond to the post-linear (order $G^2$) and post-post-linear contribution (order $G^3$) terms in the classical PM waveform \cite{Damour:2020tta,Bini:2023fiz}. An analytic expression for the waveform $W^{\alpha \beta}$ can be determined from a soft expansion~\cite{Bini:2023fiz,Georgoudis:2023eke,Georgoudis:2023ozp,Bini:2024rsy} \footnote{An apparent discrepancy between the one-loop contribution and Post-Newtonian results was noted in~\cite{Bini:2023fiz}, while certain cut contributions were argued to be missed in~\cite{Caron-Huot:2023vxl}. The resolution of these issues for the one-loop waveform result are discussed in~\cite{Georgoudis:2023eke,Georgoudis:2023ozp,Bini:2024rsy}}

\be 
W^{\alpha \beta} = \frac{1}{\omega} W^{\alpha \beta}_{\omega^{-1}} +  \ln \omega W^{\alpha \beta}_{\ln \omega} + \omega^0 W^{\alpha \beta}_{\omega^0} + \omega (\ln \omega)^2 W^{\alpha \beta}_{\ln \omega} + \omega \ln \omega W^{\alpha \beta}_{\omega \ln \omega} + \cdots
\label{wf.exp}
\ee

The pieces non-analytic in $\omega$ are constrained by classical single soft theorems~\cite{Sahoo:2021ctw,Ghosh:2021bam}, while the $\omega \ln \omega$ term additionally receives a one-loop contribution that provides a check on the one-loop corrected waveform. Our result \ref{mem.int8} could, in principle, be compared with the subleading finite frequency contribution in the soft expansion. However, this contribution (the coefficient of $\omega$) is currently absent in the literature (the $\cdots$ of \ref{wf.exp}) and involves a non-trivial computation that lies beyond the scope of our paper. 

There are known results for the linear memory effect derived from tree-level and one-loop corrected waveforms.  The leading single soft contribution to the linear memory effect from the Weinberg dressing~\cite{DiVecchia:2022nna} agrees with the tree-level contribution in~\cite{Jakobsen:2021smu}. Likewise, the leading soft limit of the one-loop corrected waveform provides the $\mathcal{O}(G^2)$ contribution to the linear memory~\cite{Herderschee:2023fxh, Georgoudis:2023eke,Bini:2024rsy}. These results are independent of the graviton frequency as a consequence of the leading single soft graviton factor. However, our result notably involves an overall dependence on the cut-off frequency for the two soft gravitons $\omega$, while also providing a non-vanishing time-independent contribution to \ref{mem.def}, which is nothing but the standard definition of memory effect. 

Hence \ref{mem.int8} in complete generality would provide a subleading contribution in the gravitational wave memory. The more general procedure involves contracting \ref{mem.int8} with polarization tensors to find a dependence on the impact parameter and impulse. As a consequence, we will find an $\mathcal{O}(G^2)$ result,  that will remain difficult to compare with the known literature. In the following, we proceed differently and consider the simplified expression in \ref{mem.lead}. The transverse-traceless components will be spatial. Using $p_n = \eta_n (E_n\,, \vec{p_n})$ and defining $\vec{v}_n = \frac{\vec{p}_n}{E_n}$, we may then express \ref{mem.lead} as
\be
\Delta W^{G^2; I J}_{\text{TT}}(x) = - \frac{8 G^2 \omega }{r} \sum_{n} \frac{m_n^2 v_n^{I} v_n^{J}}{(1 - \vec{v}_n\cdot \hat{x})^2} \,,
\label{mem.nl}
\ee
with $I\,,J$ denoting spatial indices. We will now show that this expression can be related to the nonlinear gravitational memory effect. 

We recall that the nonlinear gravitational memory is a hereditary effect purely resulting due to the gravitational emission from a scattering event with the expression~\cite{Christodoulou:1991cr,Wiseman1991ChristodoulousNG,Thorne:1992sdb,Favata:2010zu}
\be
\Delta W^{I J}_{G; \text{nonlinear}}(x) = \frac{4 G}{r}\int d\Omega \; \frac{d E^{GW}}{d \Omega} \left( \frac{\hat{n}^{I} \hat{n}^{J}}{1 - \hat{n}\cdot \hat{x}} \right)\,,
\label{gw.nlm}
\ee
with $\frac{d E^{GW}}{d \Omega}$ the differential energy distribution of the gravitational waves over the celestial sphere coordinated by $\hat{n}$ (with components $\hat{n}^{I}$) the unit spatial vector of the emitted gravitational waves, and $\hat{x}$ as before specifying the location of the detector on the celestial sphere. To relate \ref{gw.nlm} to \ref{mem.nl}, we follow in the spirit of~\cite{Thorne:1992sdb}, which relates the linear gravitational memory effect to the nonlinear one. First, we need to replace $\frac{d E^{GW}}{d \Omega}$ in \ref{gw.nlm} by the $\mathcal{O}(G)$ result of (nearly) soft gravitons described in terms of the momenta of the external particles and the graviton. Next, we need to take the ultrarelativistic limit of the massive particles so that they mimic a massless source similar to gravitons that source the nonlinear gravitational memory effect. This can be done by replacing velocity vectors $v_n^I$ of the ultrarelativistic external particles with $\hat{n}^I$, the unit spatial vector of the emitted graviton. By the above procedure, we will show that an ultrarelativistic approximation of \ref{gw.nlm} can be used to arrive at \ref{mem.nl}.

Before proceeding, we elaborate on why \ref{mem.nl} might be expected to produce a nonlinear memory effect. First, recall that the $\mathcal{O}(G^2)$ waveform \ref{mem.int7} is obtained by taking the expectation value of a single graviton as in the $\mathcal{O}(G)$ case. The contribution from the $n$ and $i$ vertices can be traced back to the process with one of the emitted gravitons from the double graviton vertex $n$ being absorbed as a Weinberg soft graviton at vertex $i$, and with the other graviton from vertex $n$ as the emitted gravitational radiation\footnote{The second line of \ref{la.int} provides the expression for this contraction, while the first term in the last line contributes to the waveform and corresponds to emission from the double graviton vertex.}. Thus,  the case with $n=i$ can be interpreted as the waveform emitted from a particle, which follows its recoil due to a soft graviton. Besides, while we cannot strictly take the massless limit of an external particle to exactly mimic a massless graviton, we may consider the ultrarelativistic limit.  In this limit, the external particle can be nearly collinear with the emitted graviton, so that we expect the resulting memory expression in \ref{mem.nl} to be proportional to the nonlinear memory effect. 

To demonstrate this, we use the 1PM expression for $E^{GW}_{\text{soft}}$ from \cite{DiVecchia:2022owy}
\be
E^{GW}_{\text{soft}} = 8 \pi G \int_{0}^{\omega} d\omega_k \int \frac{d\Omega_k}{2 (2 \pi)^3} \sum_{i,j} \frac{p_i^{\alpha} p_i^{\beta}}{E_i - \vec{p}_i\cdot \hat{k}}P_{\alpha \beta \rho \sigma} \frac{p_j^{\rho} p_j^{\sigma}}{E_j - \vec{p}_j\cdot \hat{k}}\,,
\label{egw.soft}
\ee
with $\omega$ the cut-off frequency, which in the following we do not take to vanish. Hence, the result will be more strictly for nearly soft gravitons, which are required to produce a recoil of the external particles. We accordingly consider the $i=j$ contribution in \ref{egw.soft} and take the ultrarelativistic limit by requiring $\vert\vec{v}_i \vert = 1 - \lambda$ with $\vec{v}_i := \frac{\vec{p}_i}{E_i}$ so that $m_i^2 = E_i^2 (1 - \vec{v}_i^2)\simeq 2 E_i^2 \lambda$. Here $\lambda$ represents a cut-off on the vanishing mass of the external particles. In this limit we find the following result for $\frac{d E^{GW}_{\text{soft}}}{d \Omega_k}$ from \ref{egw.soft}
\be
\frac{d E^{GW}_{\text{soft}}}{d \Omega_k} =  \frac{\omega G \lambda}{2 \pi^2} \sum_{i} m_i^2 \frac{1}{(1 - \vec{v}_i\cdot \hat{k})^2}\;.
\label{soft.ur}
\ee
We note that $\frac{d E^{GW}_{\text{soft}}}{d \Omega_k}$ is a local observable determined from the corresponding global observable in \ref{egw.soft}~\cite{Gonzo:2023cnv}. While classical global observables for gauge and gravitational theories have certain mass singularities and divergences in the ultrarelativistic limit, their corresponding local differential expressions are smooth in this limit.  The absence of any divergences in \ref{soft.ur} as $\lambda \to 0$ is consistent with this property and allows for our analysis in this limit. We refer the interested reader to~\cite{Gonzo:2023cnv} for further details and to~\cite{DiVecchia:2021ndb} for the high energy limit applied to energy spectrum and waveform derived from the coherent dressing involving Weinberg soft gravitons. For our present purpose, we note that \ref{soft.ur} provides the nearly soft energy distribution from a process involving ultrarelativistic particles with $\vert \vec{v}_i \vert = 1 - \lambda$. As these velocities are close to the speed of light, we can approximate the nonlinear memory effect in \ref{gw.nlm} by 

\be
\Delta \widetilde{W}^{I J}_{G; \text{nonlinear}}(x) \sim \frac{4 G}{r}\int d\Omega_k \sum_n \left( \frac{\omega G \lambda}{2 \pi^2} m_n^2 \frac{1}{(1 - \vec{v}_n\cdot \hat{k})^2}\right) \left( \frac{v_n^{I} v_n^{J}}{1 - \vec{v}_n\cdot \hat{k}} \right)\,.
\label{gw.nlm2}
\ee

The expression in \ref{gw.nlm2} results from \ref{gw.nlm} by replacing the energy distribution with that from \ref{soft.ur} and taking $\hat{n}^{I} \approx v^{I}_n$, and then by summing over all external particles to obtain the total contribution. Hence $\Delta \widetilde{W}^{I J}_{G; \text{nonlinear}}(x)$ in \ref{gw.nlm2} represents an ultrarelativistic particle approximation of the nonlinear memory effect expression $W^{I J}_{G; \text{nonlinear}}(x)$ in \ref{gw.nlm} sourced by nearly soft gravitons. Since we assume $\vec{v}_n$ is nearly collinear with the graviton, we can explicitly perform the integral over $\Omega_k$ by the saddle approximation, which introduces a factor of $2\pi$ from the azimuthal integral, and a condition of $\hat{k}=\hat{x}$ for the polar integral, with $\hat{x}$ as before the angular position of the detector. Hence to leading order in $\lambda$, \ref{gw.nlm2} evaluates to
\be
\Delta \widetilde{W}^{I J}_{G; \text{nonlinear}}(x) \sim \frac{2 G^2 \omega \lambda}{\pi r} \sum_{n} m_n^2 \frac{v_n^{I} v_n^{J}}{(1 - \vec{v}_n\cdot \hat{x})^2}\;.
\label{gw.nlm3}
\ee

Since the above followed from \ref{gw.nlm2}, its massless limit will agree with the nonlinear memory effect in \ref{gw.nlm}.
However, we can also establish that in the vanishing mass limit $\lambda \to 0$ of the external particles in \ref{gw.nlm3}, this nonlinear gravitational memory effect sourced by the ultrarelativistic particles can be identified with the $\mathcal{O}(G^2)$ gravitational memory effect \ref{mem.nl} due to the recoil of external particles by a nearly soft graviton through the following relation
\be 
\lim_{\lambda \to 0} \frac{4\pi}{ \lambda} \Delta \widetilde{W}^{I J}_{G; \text{nonlinear}}(x) = \Delta W^{G^2; I J}_{\text{TT}}(x)\;.
\label{relnlm}
\ee   
The overall factor of $4 \pi$ is on account of a specific angle towards the detector as opposed to an isotropic integration over angles. While this correspondence is established in a very restricted setting, it shows that the memory effect in \ref{mem.nl} can be associated with emissions from a particle following its recoil due to soft gravitons.

\subsection{Radiative momentum spectrum} \label{mom}

We will now consider the radiative momentum spectrum of soft gravitons. This starts with the classical 4-momentum of emitted soft gravitons up to ${\mathcal O}(G^2)$,
\be 
\mathcal{P}^{\alpha} = \int_{\vec{k}} d^3k \; k^{\alpha} \; \langle a^{\dagger}_i (k) a_i (k) \rangle_{\Delta_1} \;, \label{mom.def}
\ee
and the soft radiative momentum spectrum is defined by $\displaystyle{\frac{d \mathcal{P}^{\alpha}}{d \omega}}$, where $\omega$ is the cut-off frequency or the resolution power. The formal expression of $\mathcal{P}^{\alpha}$ has been given in \ref{q2.d1} with $C(k)=k^{\alpha}:=\omega_k q^{\alpha}_k$, with $q^{\alpha}_k=(1,\hat{k})$.  We first simplify 
 \ref{q2.d1} by using \ref{f1.def}, \ref{al.def}, \ref{bet.def}, and \ref{lpm} to express $\mathcal{P}^{\alpha}$ in terms of $f_{\mu\nu}(\omega_k,\hat{k})$ and $l^{\pm}_{\mu\nu}(\omega_k,\omega_l,\hat{k})$. Using the fact that the $\omega$ dependence in  $\mathcal{P}^{\alpha}$ only appears in the form of step functions such as $\Theta(\omega-\omega_k)$ or $\Theta(\omega-\omega_k-\omega_l)$, it follows that the derivative of $\mathcal{P}^{\alpha}$ with respect to $\omega$ replaces these step functions by corresponding delta functions. After these manipulations, we find 
\be
\displaystyle{\frac{d\mathcal{P}^{\alpha}}{d \omega}} = \displaystyle{\frac{d\mathcal{P}^{\alpha; G}}{d \omega}} +  \displaystyle{\frac{d\mathcal{P}^{\alpha; G^2}}{d \omega}}\;,
\ee
with
\begin{align}
\displaystyle{\frac{d\mathcal{P}^{\alpha; G}}{d \omega}} &= \frac{1}{2(2 \pi)^3}\int_{0}^{\infty} d \omega_k \omega_k^2 \oint  d\Omega_k  \,\delta\left(\omega - \omega_k \right) \, q^{\alpha}_k f^*_{\mu \nu}(\omega_k \,, \hat{k}) f^{\mu \nu}(\omega_k \,, \hat{k})\;, \label{emm.g}\\
\displaystyle{\frac{d\mathcal{P}^{\alpha; G^2}}{d \omega}} &= - \frac{1}{8(2 \pi)^4} \int_{-\infty}^{\infty} d \omega_k \omega_k^2 \int_{-\infty}^{\infty} d \omega_l \omega_l \oint  d\Omega_k  \, \Theta(\omega_k) \Theta(\omega_l) \delta\left(  \omega   -   \omega_k   -   \omega_l   \right) \, q^{\alpha}_k \notag\\
 & \qquad \qquad \qquad \left[ f^{*\,\mu \nu} (\omega_k \,, \hat{k}) \left(l^{+}_{\mu\nu}(\omega_k\,, \omega_l \,, \hat{k}) + l^{-}_{\mu\nu}(\omega_k\,, \omega_l \,, \hat{k}) \right) \right. \notag\\
& \left. \qquad \qquad \phantom{- \frac{1}{2}\int_{\vec{k}} d^3k \int_{\vec{l}} d^3l k^{\alpha}} + f^{\,\mu \nu} (\omega_k \,, \hat{k}) \left( l^{+*}_{\mu\nu}(\omega_k\,, \omega_l \,, \hat{k}) + l^{-*}_{\mu\nu}(\omega_k\,, \omega_l \,, \hat{k})\right)\right]\;.
\label{emm.g2}
\end{align}

The delta functions eliminate the subtlety of vanishing frequency; thus the integrals over $\omega_{k,l}$ can be carried out without needing a $i0$ prescription. This is different from the evaluations of waveforms and angular momentum. The integrals over $\omega_k$ and $\Omega_k$ of \ref{emm.g} have been carried out in \cite{DiVecchia:2022nna} to express $\displaystyle{\frac{d\mathcal{P}^{\alpha; G}}{d \omega}}$ in terms of kinematic variables of external particles. Here, we cite its radiative energy spectrum for later comparison to our result at ${\cal O}(G^2)$, 
\be 
\displaystyle{\frac{d E^{G}}{d \omega}}:=\displaystyle{\frac{d\mathcal{P}^{0; G}}{d \omega}} ={2 G \over \pi} \sum_{i,j}  \Big(\sigma_{ij}^2 -{1\over 2}  \Big) \eta_i \eta_j m_i m_j \Delta_{ij}\;, \label{emm.g1}
\ee
where $\Delta_{ij}$ is defined in \ref{rv.def}. In the following, we will focus on carrying out the integrals of   $\displaystyle{\frac{d \mathcal{P}^{\alpha; G^2}}{d \omega}}$ to express it in terms of the external particle kinematic variables. First, the terms appearing in the parenthesis of \ref{emm.g2} can be shown to provide
\begin{align}
&f^{*\, \mu \nu} (\omega_k \,, \hat{k}) \left(l^{+}_{\mu\nu}(\omega_k\,, \omega_l \,, \hat{k}) + l^{-}_{\mu\nu}(\omega_k\,, \omega_l \,, \hat{k})\right) = f^{\mu \nu} (\omega_k \,, \hat{k}) \left( l^{+*}_{\mu\nu}(\omega_k\,, \omega_l \,, \hat{k}) + l^{-*}_{\mu\nu}(\omega_k\,, \omega_l \,, \hat{k}) \right) \notag\\
&= 16 G^2 (2 \pi)^2 \sum_{n,i,j}  \frac{H_{nij}}{(E_n - \vec{p}_n\cdot \hat{k})(E_i - \vec{p}_i\cdot \hat{k})(E_j - \vec{p}_j\cdot \hat{k})}\frac{1}{\omega_k \omega_l}\left( \frac{1}{\omega_k -\omega_l} + \frac{1}{\omega_k + \omega_l} \right)\,,
\label{la2.int}
\end{align}
with
\be  
H_{nij}:= \eta_n \eta_i \eta_j m_n^2 m_i^2 m_j^2 \left(1 + 4 \sigma_{ij}\sigma_{ni}\sigma_{nj} - 2 \sigma_{ni}^2 - 2 \sigma_{nj}^2 \right) \;. \label{hnim.def}
\ee

We note that the function $H_{nij}$ of the relative velocities for the external particles is symmetric in $i$ and $j$, i.e., $H_{nij} = H_{nji}$. The $i$ and $j$ indices are particle labels on which the Weinberg soft gravitons are attached. However, there is no symmetry of $H_{nij}$ associated with the particle label $n$ for the double graviton vertex. This reflects an underlying feature in the kinematic factors for the momentum and angular momentum results.

Substituting \ref{la2.int} in \ref{emm.g2}, we find a frequency integral which can be evaluated to yield a simple result,
\begin{align}
& \int_{-\infty}^{\infty} d \omega_k \int_{-\infty}^{\infty} d \omega_l  \, \Theta(\omega_k) \Theta(\omega_l)\delta\left(\omega - \omega_k - \omega_l \right)  \left( \frac{\omega_k}{\omega_k -\omega_l} + \frac{\omega_k}{\omega_k + \omega_l} \right)\;, \notag\\
&= \int_{0}^{\omega} d \omega_k \frac{\omega_k}{\omega} + \frac{1}{4}\int_{-\omega}^{\omega} d\omega_k \left(1 + \frac{\omega}{\omega_k}\right) = \omega\;.\label{ang.omega.f}
\end{align}

Combining the results of \ref{la2.int} and \ref{ang.omega.f}, we can reduce \ref{emm.g2} to
\begin{align}
\frac{d \mathcal{P}^{\alpha; G^2}}{d \omega} &= - \frac{2 G^2 \omega}{\pi} \sum_{n,i,j} H_{nij} \; \big(F_{nij}, \vec{G}_{nij} \big) \;,
\label{emm.g21}
\end{align}
where the $4$-vector notation $\big(F_{nij}, \vec{G}_{nij} \big)$ indicates that its temporal component $F_{nij}$ determines $\frac{d \mathcal{P}^{0; G^2}}{d \omega}$ and hence the radiated energy spectrum, while the spatial $3$-vector $\vec{G}_{nij}$ determines the radiated spatial momentum, and these terms have the definitions
\begin{align}
F_{nij} & := {1\over 2\pi} \oint d\Omega_k  \frac{1}{(E_n - \vec{p}_n\cdot \hat{k})(E_i - \vec{p}_i\cdot \hat{k})(E_j - \vec{p}_j\cdot \hat{k})}  \,,\label{ang.f}\\ 
\vec{G}_{nij} &:= {1\over 2\pi} \oint d\Omega_k  \frac{\hat{k}}{(E_n - \vec{p}_n\cdot \hat{k})(E_i - \vec{p}_i\cdot \hat{k})(E_j - \vec{p}_j\cdot \hat{k})} \,
\label{ang.g}
\end{align}

The angular integrals of \ref{ang.f} and \ref{ang.g} have been carried out in Appendix \ref{B}. Both $F_{nij}$ and $\vec{G}_{nij}$ are completely symmetric under the interchange of the external particle labels $n, i$ and $j$. While the expression for $\vec{G}_{nij}$ is quite involved, $F_{nij}$ is considerably simpler with the result,
\begin{align}
F_{nij} & = \frac{2\; (m_n\Delta_{ij}+ m_i \Delta_{nj}+m_j \Delta_{ni}) }{m_n m_i m_j\sqrt{2 (m_i m_j \sigma_{ij}+ m_n m_j \sigma_{nj} + m_n m_i \sigma_{ni}) + m_n^2  + m_i^2 + m_j^2}} \;.
\label{fnim.def}
\end{align}

Thus, using \ref{hnim.def} and \ref{fnim.def} we can obtain the explicit expression for the soft radiative energy spectrum at ${\cal O}(G^2)$  
\begin{align}
& \frac{dE^{G^2}}{d \omega} :=\displaystyle{\frac{d\mathcal{P}^{0; G^2}}{d \omega}} =  -\frac{2 G^2 \omega}{\pi} \sum_{n,i,j}  H_{nij} F_{nij}\;.
\label{egw.res2}
\end{align}

This result can be directly related to the $\kappa^4$ correction of $\text{Im} 2 \delta$. Using the definitions in \ref{f1.def}, \ref{al.def}, \ref{bet.def}, and subsequently \ref{hnim.def} and \ref{ang.f} in \ref{im2d.k3} we find
\be
\text{Im} 2\delta_{\kappa^4} = -\frac{2 G^2}{\pi \hbar} \int_{0}^{\omega} d \omega_k \int_{0}^{\omega - \omega_k} d\omega_l \frac{\omega_k}{ \omega_k^2 - \omega_l^2} \sum_{n,i,j} H_{nij} F_{nij}\,,
\label{im2dk3.ev}
\ee
where we made use of the step function in the limits of the $\omega_k$ and $\omega_l$ integrals, and the integrand follows from the absence of any $i0$ poles being considered in the derivation (as with the energy spectrum). We now note that the $\omega_k$ and $\omega_l$ integrals do not involve any divergences and can be evaluated without dimensional regularization. We have
\be
\int_{0}^{\omega} d \omega_k \int_{0}^{\omega - \omega_k} d \omega_l  \frac{\omega_k}{ \omega_k^2 - \omega_l^2}   = \frac{\omega}{2}\;,
\label{om.int}
\ee
and hence using \ref{om.int} in \ref{im2dk3.ev} gives us the result
\be
\text{Im} 2\delta_{\kappa^4} = -\frac{G^2 \omega}{\pi \hbar} \sum_{n,i,j} H_{nij} F_{nij}\,.
\label{im2dk3.ans}
\ee

Comparing \ref{im2dk3.ans} with \ref{egw.res2} gives the relation
\be \frac{\hbar}{2} \text{Im} 2\delta_{\kappa^4} = \frac{dE^{G^2}}{d \omega}\,,  \label{imew}
\ee
This provides a generalization to ${\cal O}(G^2)$ of the relationship between the imaginary part of the eikonal and the emitted energy spectrum in~\cite{DiVecchia:2021ndb, DiVecchia:2022nna} in the infrared limit, which we recall here for comparison,
\be
\lim_{\epsilon\rightarrow 0} \big[ -4\hbar \epsilon {\rm Im} 2 \delta_{\kappa^2}  \big] = \lim_{\omega\rightarrow 0} \frac{dE^G}{d \omega} \label{1Grr}
\ee
with $\epsilon = \frac{4 - d}{2}$ the dimensional regularization parameter. While the infrared diverges in $\text{Im} 2\delta_{\kappa^2}$ are related to the zero frequency limit of $\frac{dE^{G}}{d \omega}$ as shown in \ref{1Grr}, we note that \ref{imew} holds for all subleading $\omega$. More specifically, \ref{imew} provides a $\mathcal{O}(G^2)$ and higher relation between a subleading soft radiation energy spectrum and the imaginary part of the eikonal phase from next-to-eikonal corrected soft factors.

\subsection{Radiative angular momentum} \label{amom}

We lastly consider the case of angular momentum $J^{\alpha \beta} = L^{\alpha \beta} + S^{\alpha \beta}$,  where $L^{\alpha \beta}$ and $S^{\alpha \beta}$ respectively denote the orbital and spin angular momentum and have the following operator expression in the de Donder gauge,
\begin{align}
 L^{\alpha \beta} &= - i \eta^{\alpha \gamma} \eta^{\beta \delta} P^{\mu \nu \rho \sigma} \int_{\vec{k}} d^3k  \,a^{\dagger}_i (k) \varepsilon^*_{i\,,\mu \nu} (k) k_{[\gamma} \frac{\overset{\leftrightarrow}{\partial}}{\partial k^{\delta]}}
 a_j (k) \varepsilon_{j\,,\rho \sigma} (k)\;,  \label{omom.def}\\
S^{\alpha \beta} &= - 2 i \eta^{\alpha \gamma} \eta^{\beta \delta} \eta^{\mu \rho} \delta^{\nu}_{[\gamma} \delta^{\sigma}_{\delta]} \int_{\vec{k}} d^3k \, a^{\dagger}_i (k) a_j (k) \varepsilon^*_{i\,,\mu \nu} (k) \varepsilon_{j\,,\rho \sigma} (k) \;, \label{spin.def}
\end{align}
with $P^{\mu \nu \rho \sigma}$ given in \ref{Pp.def}. Our symmetrization notations and conventions for the left-right derivative are 
\begin{align}
A_{[\gamma}B_{\delta]} = A_{\gamma} B_{\delta} - A_{\delta} B_{\gamma} \,, &\quad  A_{(\gamma}B_{\delta)} = A_{\gamma} B_{\delta} + A_{\delta} B_{\gamma}\,, \notag\\
 A k_{\gamma} \frac{\overset{\leftrightarrow}{\partial}}{\partial k^{\delta}}B = \frac{1}{2} &\left(A k_{\gamma}\frac{\partial B}{\partial k^{\delta}} - k_{\gamma}\frac{\partial A}{\partial k^{\delta}} B  \right)\;.
\end{align}

In the following, we consider the orbital and spin contributions separately to briefly discuss a gauge invariance issue. We define the expectation value with respect to the dressed states as $\mathcal{L}^{\alpha \beta}:= \langle L^{\alpha \beta} \rangle_{\Delta}$ and $\mathcal{S}^{\alpha \beta}:= \langle S^{\alpha \beta} \rangle_{\Delta}$. We further expand the expectation values in terms of their $G$  and $G^2$ contributions, as denoted by $\mathcal{L}^{\alpha \beta} = \mathcal{L}^{\alpha \beta;G} + \mathcal{L}^{\alpha \beta;G^2} + \mathcal{O} (G^3)$ and $\mathcal{S}^{\alpha \beta} = \mathcal{S}^{\alpha \beta;G} + \mathcal{S}^{\alpha \beta;G^2} + \mathcal{O} (G^3)$, and after straightforward manipulations we find 
\begin{align}
\mathcal{L}^{\alpha \beta;G} & = -i \eta^{\alpha \gamma} \eta^{\beta \delta} P^{\mu \nu \rho \sigma}\frac{1}{2 (2 \pi)^3} \int_{0}^{\omega} d \omega_k \omega_k \oint  d\Omega_k \,f^*_{\mu \nu}(\omega_k\,, \hat{k}) k_{[\gamma} \frac{\overset{\leftrightarrow}{\partial}}{\partial k^{\delta]}}
 f_{\rho \sigma}(\omega_k\,, \hat{k})\;, \label{oamG}\\
\mathcal{L}^{\alpha \beta;G^2} & = i \eta^{\alpha \gamma} \eta^{\beta \delta} P^{\mu \nu \rho \sigma} \frac{1}{8 (2 \pi)^4} \int_{0}^{\omega} d \omega_k \omega_k  \oint  d\Omega_k  \,   \notag\\
 & \,\qquad   \left[   \int_{0}^{\omega - \vert \omega_k\vert} d \omega_l \omega_l \left( l^{+*}_{\mu\nu}(\omega_k\,, \omega_l \,, \hat{k}) + l^{-*}_{\mu\nu}(\omega_k\,, \omega_l \,, \hat{k}) \right) k_{[\gamma} \frac{\overset{\leftrightarrow}{\partial}}{\partial k^{\delta]}} f_{\rho \sigma}(\omega_k\,, \hat{k}) \right. \notag\\
 &\left. \qquad  \qquad   + f^*_{\mu \nu}(\omega_k\,, \hat{k}) k_{[\gamma}\frac{\overset{\leftrightarrow}{\partial}}{\partial k^{\delta]}} \int_{0}^{\omega - \vert \omega_k\vert} d \omega_l \omega_l \left( l^{+}_{\rho \sigma}(\omega_k\,, \omega_l \,, \hat{k}) + l^{-}_{\rho \sigma}(\omega_k\,, \omega_l \,, \hat{k}) \right) \right]\label{oamG2.1}
 \end{align}
for the orbital angular momentum and 
\begin{align}
\mathcal{S}^{\alpha \beta;G} & = - i \eta^{\alpha \gamma} \eta^{\beta \delta} \eta^{\mu \rho} \delta^{\nu}_{[\gamma} \delta^{\sigma}_{\delta]}\frac{1}{(2 \pi)^3} \int_{0}^{\omega} d \omega_k \omega_k  \oint  d\Omega_k   \,f^*_{\mu \nu}(\omega_k\,, \hat{k}) f_{\rho \sigma}(\omega_k\,, \hat{k})\;,  \label{samG}\\
 \mathcal{S}^{\alpha \beta;G^2} & = i \eta^{\alpha \gamma} \eta^{\beta \delta} \eta^{\mu \rho} \delta^{\nu}_{[\gamma} \delta^{\sigma}_{\delta]} \frac{1}{4 (2 \pi)^4} \int_{0}^{\omega} d \omega_k \omega_k  \oint  d\Omega_k \, \notag\\
 & \,\qquad  \left[\int_{0}^{\omega - \vert \omega_k\vert} d \omega_l \omega_l \left( l^{+*}_{\mu\nu}(\omega_k\,, \omega_l \,, \hat{k}) + l^{-*}_{\mu\nu}(\omega_k\,, \omega_l \,, \hat{k})\right) f_{\rho \sigma}(\omega_k\,, \hat{k}) \right. \notag\\
 &\left. \qquad \qquad + f^*_{\mu \nu}(\omega_k\,, \hat{k}) \int_{0}^{\omega - \vert \omega_k\vert} d \omega_l \omega_l \left( l^{+}_{\rho\sigma}(\omega_k\,, \omega_l \,, \hat{k}) + l^{-}_{\rho \sigma}(\omega_k\,, \omega_l \,, \hat{k})\right) \right]\label{samG2.1}
 \end{align}
 for the spin contribution. The step functions restricting the sum of $\omega_k$ and $\omega_l$ to be bounded by $\omega$ have been considered explicitly in the integration limits of \ref{oamG2.1} and \ref{samG2.1}. This will be particularly useful in considering the action of the operator $k_{[\alpha} \frac{\partial}{\partial k^{\beta]}}$ in \ref{oamG2.1}. 

The ${\cal O}(G)$ terms \ref{oamG} and \ref{samG} have been considered and evaluated in \cite{DiVecchia:2022owy}. These two terms are separately gauge non-invariant due to contributions from the projection operator of \ref{proj.d} hidden in $f_{\mu\nu}$. However, the gauge non-invariant pieces cancel out and the resultant gauge invariant sum gives \cite{DiVecchia:2022owy}
\be
 \mathcal{J}^{\alpha \beta;G} :=\mathcal{L}^{\alpha \beta;G}+ \mathcal{S}^{\alpha \beta;G} ={G\over 2}\sum_{i,j}\left[\left(\sigma^2_{ij}-\frac{1}{2} \right){\sigma_{ij}\Delta_{ij}-1 \over \sigma^2_{ij}-1} - 2 \sigma_{ij}\Delta_{ij}\right](\eta_i-\eta_j) p_i^{[\alpha}p_j^{\beta]}\;.  
\ee

To discuss the gauge issue further for the  ${\cal O}(G^2)$ terms \ref{oamG2.1} and \ref{samG2.1}, we make explicit the gauge vector $\lambda^{\mu}$ terms from the projection operator of \ref{proj.d} and \ref{proj.p} in the the expressions of $f_{\rho \sigma}(\omega_k\,, \hat{k})$ and $l^{\pm}_{\rho\sigma}(\omega_k\,, \omega_l \,, \hat{k})$ as follows:
\begin{equation}
 f_{\rho \sigma}(\omega_k\,, \hat{k}) = F_{\rho \sigma}(\omega_k\,, \hat{k}) + \lambda^{\alpha}F_{\alpha (\rho}(\omega_k\,, \hat{k}) k_{\sigma)} + \lambda^{\alpha} F_{\alpha \beta}(\omega_k\,, \hat{k}) \lambda^{\beta} k_{\rho} k_{\sigma} - \eta^{\alpha \beta} F_{\alpha \beta}(\omega_k\,, \hat{k}) \Pi_{\rho \sigma}(\hat{k})\;,
\label{fp}
\end{equation}
and
\begin{align}
l^{\pm}_{\rho\sigma}(\omega_k\,, \omega_l \,, \hat{k})  &= C^{\pm}_{\rho \sigma}(\omega_k\,, \omega_l \,, \hat{k}) + \lambda^{\alpha}C^{\pm}_{\alpha (\rho}(\omega_k\,, \omega_l \,, \hat{k}) k_{\sigma)} \notag\\
& \qquad \qquad + \lambda^{\alpha} C^{\pm}_{\alpha \beta}(\omega_k\,, \omega_l \,, \hat{k}) \lambda^{\beta} k_{\rho} k_{\sigma} - \eta^{\alpha \beta} C^{\pm}_{\alpha \beta}(\omega_k\,, \omega_l \,, \hat{k}) \Pi_{\rho \sigma}(\hat{k})\;.
\label{lp}
\end{align}
Plugging the above into \ref{oamG2.1} and \ref{samG2.1}, we can single out the $\lambda^{\mu}$-dependent and hence the gauge non-invariant pieces to find
\begin{align}
 \mathcal{L}^{\alpha \beta;G^2} & =  \tilde{\mathcal{L}}^{\alpha \beta; G^2} - \mathcal{G}^{\alpha \beta; G^2} \;, \label{oamG2.2}  \\
 \mathcal{S}^{\alpha \beta;G^2} & =  \tilde{\mathcal{S}}^{\alpha \beta; G^2} + \mathcal{G}^{\alpha \beta; G^2}\;, \label{samG2.2}
\end{align}
where the gauge invariant ($\lambda^{\mu}$-independent) pieces are given by 
\begin{align}
\tilde{\mathcal{L}}^{\alpha \beta; G^2}  &= \frac{i}{8 (2 \pi)^4} \oint  d\Omega_k \eta^{\alpha \gamma} \eta^{\beta \delta} \int_{0}^{\omega} d \omega_k \,\omega_k   \notag\\
&\quad \, \left[\int_{0}^{\omega - \vert \omega_k\vert} d \omega_l \omega_l  \left( C^{+*}_{\mu\nu}(\omega_k\,, \omega_l \,, \hat{k}) + C^{-*}_{\mu\nu}(\omega_k\,, \omega_l \,, \hat{k}) \right) P^{\mu \nu \rho \sigma} k_{[\gamma} \frac{\overset{\leftrightarrow}{\partial}}{\partial k^{\delta]}} F_{\rho \sigma}(\omega_k\,, \hat{k}) \right. \notag\\
&\left. \quad \; + F^*_{\mu \nu}(\omega_k\,, \hat{k}) P^{\mu \nu \rho \sigma} k_{[\gamma} \frac{\overset{\leftrightarrow}{\partial}}{\partial k^{\delta]}} \int_{0}^{\omega - \vert \omega_k\vert} d \omega_l \omega_l  \left( C^{+}_{\rho \sigma}(\omega_k\,, \omega_l \,, \hat{k}) + C^{-}_{\rho\sigma}(\omega_k\,, \omega_l \,, \hat{k}) \right) \right] \,, \label{oam.gi}\\
 \tilde{\mathcal{S}}^{\alpha \beta; G^2}  &= \frac{i}{4 (2 \pi)^4} \oint  d\Omega_k \eta^{\alpha \gamma} \eta^{\beta \delta} \int_{0}^{\omega} d \omega_k \, \omega_k   \notag\\
&\, \quad \left[\int_{0}^{\omega - \vert \omega_k\vert} d \omega_l \omega_l  \left( C^{+*}_{\mu\nu}(\omega_k\,, \omega_l \,, \hat{k}) + C^{-*}_{\mu\nu}(\omega_k\,, \omega_l \,, \hat{k}) \right) \eta^{\mu \rho} \delta^{\nu}_{[\gamma} \delta^{\sigma}_{\delta]} F_{\rho \sigma}(\omega_k\,, \hat{k}) \right. \notag\\
&\left. \qquad \; + F^*_{\mu \nu}(\omega_k\,, \hat{k}) \eta^{\mu \rho} \delta^{\nu}_{[\gamma} \delta^{\sigma}_{\delta]} \int_{0}^{\omega - \vert \omega_k\vert} d \omega_l \omega_l  \left( C^{+}_{\rho \sigma}(\omega_k\,, \omega_l \,, \hat{k}) + C^{-}_{\rho\sigma}(\omega_k\,, \omega_l \,, \hat{k}) \right) \right] \,, \label{sam.gi}
\end{align}
and the common gauge non-invariant 
($\lambda^{\mu}$-dependent) piece is given by 
\begin{align}
 \mathcal{G}^{\alpha \beta; G^2} = \frac{i}{4 (2 \pi)^4}\int_{0}^{\omega} d \omega_k \omega_k  \oint  d\Omega_k \eta^{\alpha \gamma} \eta^{\beta \delta} \int_{0}^{\omega - \vert \omega_k\vert} d \omega_l \omega_l   k_{[\gamma} \mathcal{E}_{\delta]}  
\end{align}
with 
\begin{align}
& \mathcal{E}_{\beta} = \lambda^{\mu} \eta^{\nu \gamma} \left(\left( C^{+*}_{\mu\nu}(\omega_k\,, \omega_l \,, \hat{k}) + C^{-*}_{\mu\nu}(\omega_k\,, \omega_l \,, \hat{k}) \right)F_{\gamma \beta} - F_{\mu \nu}\left( C^{+*}_{\gamma \beta}(\omega_k\,, \omega_l \,, \hat{k}) + C^{-*}_{\gamma \beta}(\omega_k\,, \omega_l \,, \hat{k}) \right) \right. \notag\\
&\left.\qquad  -\left( C^{+}_{\mu\nu}(\omega_k\,, \omega_l \,, \hat{k}) + C^{-}_{\mu\nu}(\omega_k\,, \omega_l \,, \hat{k}) \right)F^*_{\gamma \beta} + F^*_{\mu \nu}\left( C^{+}_{\gamma \beta}(\omega_k\,, \omega_l \,, \hat{k}) + C^{-}_{\gamma \beta}(\omega_k\,, \omega_l \,, \hat{k}) \right) \right)\;.
 \end{align}

We now see that the gauge non-invariant pieces in \ref{oamG2.2} and \ref{samG2.2} cancel out, and the total angular momentum $\mathcal{J}^{\alpha \beta;G^2} =  \tilde{\mathcal{L}}^{\alpha \beta; G^2} + \tilde{\mathcal{S}}^{\alpha \beta; G^2}$  at  ${\cal O}(G^2)$ is gauge invariant and with the general form in~\cite{Manohar:2022dea}.

Making the transformation   $\omega_k \to -\omega_k$ and $\omega_l \to -\omega_l$ in the integrands of \ref{oam.gi} and \ref{sam.gi}, and using the properties in \ref{fab.id} with the following identity
\begin{align}
\int_{0}^{\omega - \vert \omega_k\vert} d \omega_l \omega_l & \left( C^{+}_{\rho \sigma}(\omega_k\,, \omega_l \,, \hat{k}) + C^{-}_{\rho\sigma}(\omega_k\,, \omega_l \,, \hat{k}) \right) = \int_{-\omega + \vert \omega_k\vert}^{\omega - \vert \omega_k\vert} d \omega_l \omega_l C^{+}_{\rho \sigma}(\omega_k\,, \omega_l \,, \hat{k}) \notag\\
& = \int_{-\omega + \vert \omega_k\vert}^{0} d \omega_l \omega_l  \left( C^{+}_{\rho \sigma}(\omega_k\,, \omega_l \,, \hat{k}) + C^{-}_{\rho\sigma}(\omega_k\,, \omega_l \,, \hat{k}) \right) \,,
\label{cint}
\end{align}
to make the integral over $\omega_l$ symmetric, we obtain
\begin{align}
\tilde{\mathcal{L}}^{\alpha \beta; G^2}  &= \frac{i}{8 (2 \pi)^4} \oint  d\Omega_k \eta^{\alpha \gamma} \eta^{\beta \delta}  P^{\mu \nu \rho \sigma}  \int_{-\omega}^{\omega} d \omega_k \,\omega_k  F^*_{\mu \nu}(\omega_k\,, \hat{k}) k_{[\gamma} \frac{\overset{\leftrightarrow}{\partial}}{\partial k^{\delta]}} \int_{-\omega + \vert \omega_k\vert}^{\omega - \vert \omega_k\vert} d \omega_l \omega_l   C^{+}_{\rho \sigma}(\omega_k\,, \omega_l \,, \hat{k})\;, \label{oam.gi3}\\
 \tilde{\mathcal{S}}^{\alpha \beta; G^2}  &= \frac{i}{4 (2 \pi)^4} \oint  d\Omega_k \eta^{\alpha \gamma} \eta^{\beta \delta} \eta^{\mu \rho} \delta^{\nu}_{[\gamma} \delta^{\sigma}_{\delta]} \int_{-\omega}^{\omega} d \omega_k \, \omega_k \int_{-\omega + \vert \omega_k\vert}^{\omega - \vert \omega_k\vert} d \omega_l \omega_l F^*_{\mu \nu}(\omega_k\,, \hat{k}) C^{+}_{\rho \sigma}(\omega_k\,, \omega_l \,, \hat{k})  \,. \label{sam.gi3}
\end{align}

Using \ref{Cpm.def}, \ref{pvd} and \ref{ang.f}, we can carry out the integrations of \ref{sam.gi3} and obtain the final expression for the gauge-invariant contribution to the spin angular momentum, 
\begin{align}
& \tilde{\mathcal{S}}^{\alpha \beta; G^2}  =  4 G^2 \omega \sum_{n,i,j} m_n m_i m_j^2 \eta_j (\eta_i - \eta_n) F_{nij} p_i^{[\alpha} p_n^{\beta]} (\sigma_{nj} \sigma_{ij} - \sigma_{ni})
\label{sam.fin}
\end{align}
where $F_{nij}$ is given by \ref{fnim.def}. In deriving \ref{sam.fin}, we made use of the antisymmetry in the particle labels within the sum due to $p_i^{[\alpha} p_n^{\beta]} = p_{[i}^{\alpha} p_{n]}^{\beta}$. 

For the orbital angular momentum, we first note that from the relation $k^{\mu} = \omega_k q^{\mu}_k$ in \ref{mom.param} that
\begin{equation}
    k_{[\gamma} \frac{\partial}{\partial k^{\delta]}} =   q_{k [\gamma} \frac{\partial}{\partial q_k^{\delta]}} -  q_{k [\gamma} \bar{q}_{k \delta]} \left( q_k^{\alpha} \frac{\partial}{\partial q_k^{\alpha}} - \omega_k \frac{\partial}{\partial \omega_k} \right)
    \label{par.d}
\end{equation}
with
\begin{equation}
q_k^{\alpha} = (1\,, \hat{k}) \;; \qquad \bar{q}_k^{\alpha} = \frac{1}{2}(-1\,, \hat{k}) \;; \qquad \eta_{\alpha \beta} q_k^{\alpha} \bar{q}_k^{\beta} = 1\;.
\end{equation}
Hence \ref{par.d} can be expressed as
\begin{equation}
    k_{[\gamma} \frac{\partial}{\partial k^{\delta]}} =   q_{k [\gamma} \frac{\partial}{\partial q_k^{\delta]}} +  \delta_{0 [\gamma} \delta_{ \delta]I} \hat{k}^{I} \left( q_k^{\alpha} \frac{\partial}{\partial q_k^{\alpha}} - \omega_k \frac{\partial}{\partial \omega_k} \right)
    \label{par}
\end{equation}
Note that the second term on the R.H.S. of \ref{par} satisfies 
\be \left(q_k^{\alpha} \frac{\partial}{\partial q_k^{\alpha}} - \omega_k \frac{\partial}{\partial \omega_k} \right) \frac{1}{\omega_k p\cdot q_k} = 0\;, \label{scal.op} \ee
which can be used to simplify the momentum-derivative in \ref{oam.gi3} as follows
\begin{align}
F^*_{\mu \nu}(\omega_k\,, \hat{k}) k_{[\gamma} \frac{\overset{\leftrightarrow}{\partial}}{\partial k^{\delta]}}    C^{+}_{\rho \sigma}(\omega_k\,, \omega_l \,, \hat{k}) &=  \frac{1}{2} F^*_{\mu \nu}(\omega_k\,, \hat{k})  \delta_{0 [\gamma} \delta_{ \delta]I} \hat{k}^{I} \left( q_k^{\alpha} \frac{\partial}{\partial q_k^{\alpha}} - \omega_k \frac{\partial}{\partial \omega_k} \right) C^{+}_{\rho \sigma}(\omega_k\,, \omega_l \,, \hat{k})
 \notag \\
 & \quad  + F^*_{\mu \nu}(\omega_k\,, \hat{k}) q_{[\gamma} \frac{\overset{\leftrightarrow}{\partial}}{\partial q^{\delta]}}    C^{+}_{\rho \sigma}(\omega_k\,, \omega_l \,, \hat{k}) \;. \label{part.1}
\end{align}
This separation gives the following decomposition 
\be\tilde{\mathcal{L}}^{\alpha \beta; G^2} = \tilde{\mathcal{L}}_1^{\alpha \beta; G^2} + \tilde{\mathcal{L}}_2^{\alpha \beta; G^2}
\ee
with 
\begin{align}
&\tilde{\mathcal{L}}_1^{\alpha \beta; G^2} =  
-\frac{ i G^2}{(2 \pi)^2} \sum_{n,i,j} H_{nij} \oint  d\Omega_k \delta_{0}^{ [\alpha} \delta^{\beta]}_{I} \hat{k}^{I}  \int_{-\omega}^{\omega} d \omega_k  \; \omega_k \frac{1}{\omega_k - i \eta_i 0_k} \frac{1}{\eta_i p_i.q_k}  \notag\\
& \qquad \, \left( q_k^{\gamma} \frac{\partial}{\partial q_k^{\gamma}} - \omega_k \frac{\partial}{\partial \omega_k} \right) \int_{-\omega + \vert \omega_k\vert}^{\omega - \vert \omega_k\vert} d \omega_l \frac{1}{ \omega_k + \omega_l + i \eta_n 0_k + i \eta_n 0_l} \frac{1}{(\eta_n p_n\cdot q_k) (\eta_j p_j\cdot q_k)}\;, \label{l1.def}\\
&\tilde{\mathcal{L}}_2^{\alpha \beta; G^2} =  
-\frac{2 i G^2}{(2 \pi)^2} \sum_{n,i,j} H_{nij} \eta^{\alpha \gamma} \eta^{\beta \delta} \oint  d\Omega_k  \int_{-\omega}^{\omega} d \omega_k \omega_k \; \frac{1}{\omega_k - i \eta_i 0_k} \frac{1}{\eta_i p_i\cdot q_k} \notag\\
& \qquad  q_{k[\gamma} \frac{\overset{\leftrightarrow}{\partial}}{\partial q_k^{\delta]}} \int_{-\omega + \vert \omega_k\vert}^{\omega - \vert \omega_k\vert} d \omega_l \frac{1}{ \omega_k + \omega_l + i \eta_n 0_k + i \eta_n 0_l} \frac{1}{(\eta_n p_n\cdot q_k) (\eta_j p_j\cdot q_k)} \;. \label{l2.def}
\end{align}
To arrive at the above, we again used \ref{f1.def}, \ref{Cpm.def} and \ref{hnim.def} for simplification.

Performing the momentum derivatives in \ref{l1.def}, and then the $\omega_k$ integral by principal value identity \ref{pvd}, and using the defining equation of $\vec{G}_{nij}$, we arrive at the final expression of $\tilde{\mathcal{L}}_1^{\alpha \beta; G^2}$, i.e.,  
\begin{align}
\tilde{\mathcal{L}}_1^{\alpha \beta; G^2} =  
- 2 G^2 \omega \sum_{n,i,j} \eta_n H_{nij}\delta_{0}^{ [\alpha} \delta^{\beta]}_{I} G^I_{nij} \label{l1.fin}
\end{align}
where $G^I_{nij}:=(\vec{G}_{nij})^I$ of \ref{ang.g} with its expression given in Appendix \ref{Appendix.G}. We note that \ref{l1.fin} represents a mass-dipole contribution to the angular momentum \cite{DiVecchia:2022owy,Manohar:2022dea}. From the derivation, \ref{l1.fin} is due to the different energies of the two collinear gravitons from the double graviton vertex.

We finally come to the second piece of orbital angular momentum \ref{l2.def}. Using \ref{f1.def}, \ref{Cpm.def} and \ref{hnim.def}, and carrying out the integrals over $\omega_k$ and $\omega_l$ as above, we can simplify it to the following: 
\begin{align}
&\tilde{\mathcal{L}}_2^{\alpha \beta; G^2} =  
G^2 \omega \sum_{n,i,j} \eta_n H_{nij}\left[ - K^{[\alpha}_i p_i^{\beta]} + K^{[\alpha}_n p_n^{\beta]} + K^{[\alpha}_j p_j^{\beta]}\right]  \label{l2.1}
\end{align}
with 
\begin{align}
K^{\alpha}_a = {1\over 2\pi} \oint  d\Omega_k \frac{\eta_a q^{\alpha}_k}{(\eta_a p_a\cdot q_k)^2 (\eta_b p_b\cdot q_k)(\eta_c p_c\cdot q_k)} \qquad  \mbox{for} \quad a \neq b \neq c\;.
\label{K.int}
\end{align}

We can carry out the integral \ref{K.int} along similar lines of \cite{DiVecchia:2022owy}. This follows from \ref{K.int} being a Lorentz vector and hence having the general solution
\begin{equation}
K^{\alpha}_a =  A_a^{n} p_{n}^{\alpha} + A_a^{i} p_{i}^{\alpha}  + A_a^{j} p_{j}^{\alpha}\;.
\label{K.lor}
\end{equation}
The expressions for the coefficients in \ref{l2.1} can be readily derived from \ref{K.lor} 
\begin{align}
A_a^{i} & = \frac{K_a\cdot p_i \frac{m_n m_j}{m_i} (\sigma_{nj}^2 - 1) - \eta_i \left(K_a\cdot p_n \eta_n m_j (\sigma_{ij}\sigma_{nj} - \sigma_{ni}) + K_a\cdot p_j \eta_j m_n (\sigma_{ni}\sigma_{nj} - \sigma_{ij})  \right)}{m_n m_i m_j (1 + 2 \sigma_{ij}\sigma_{ni}\sigma_{nj} - \sigma_{ij}^2 - \sigma_{ni}^2- \sigma_{nj}^2)}\;, \notag\\
A_a^{n} & = \frac{K_a\cdot p_n \frac{m_i m_j}{m_n} (\sigma_{ij}^2 - 1) - \eta_n \left(K_a\cdot p_i \eta_i m_j (\sigma_{nj}\sigma_{ij} - \sigma_{ni}) + K_a\cdot p_j \eta_j m_i (\sigma_{in}\sigma_{ij} - \sigma_{nj})  \right)}{m_n m_i m_j (1 + 2 \sigma_{ij}\sigma_{ni}\sigma_{nj} - \sigma_{ij}^2 - \sigma_{ni}^2- \sigma_{nj}^2)}\;, \notag\\
A_a^{j} & = \frac{K_a\cdot p_j \frac{m_n m_i}{m_j} (\sigma_{ni}^2 - 1) - \eta_j \left(K_a\cdot p_i \eta_i m_n (\sigma_{nj}\sigma_{ni} - \sigma_{ij}) + K_a\cdot p_n \eta_n m_i (\sigma_{ij}\sigma_{ni} - \sigma_{nj})  \right)}{m_n m_i m_j (1 + 2 \sigma_{ij}\sigma_{ni}\sigma_{nj} - \sigma_{ij}^2 - \sigma_{ni}^2- \sigma_{nj}^2)}\;.
\label{Ac.def}
\end{align}
Each $K_a \cdot p$ is a special case of the integral in \ref{ang.f}. From \ref{K.int} and using $\eta_a p_a \cdot q_k = - (E_a - \vec{p}_a \cdot \hat{k})$ we have
\begin{align}
K_a\cdot p_a &= 
- {1\over 2\pi}  \oint d\Omega_k  \frac{1}{(E_n - \vec{p}_n\cdot \hat{k})(E_i - \vec{p}_i\cdot\hat{k})(E_j - \vec{p}_j\cdot \hat{k})} = - F_{nij}\;, \notag\\
K_a\cdot p_b & = - {\eta_a \eta_b \over 2\pi}  \oint d\Omega_k  \frac{1}{(E_a - \vec{p}_a\cdot \hat{k})^2 (E_c - \vec{p}_c\cdot\hat{k})} = - \eta_a \eta_b F_{aac}\,,
\label{kp.gen}
\end{align}
which provides the following relations
\begin{align}
&K_n\cdot p_n = K_i\cdot p_i =  K_j\cdot p_j = - F_{nij}\;; \notag\\
K_n\cdot p_i = - \eta_n \eta_i F_{nnj} \;; \qquad & K_i\cdot p_j = - \eta_i \eta_j F_{iin} \;; \qquad  K_i\cdot p_n = - \eta_i \eta_n F_{iij}\;; \notag\\
K_n\cdot p_j = - \eta_n \eta_j F_{nni} \;; \qquad & K_j\cdot p_i = - \eta_i \eta_j F_{jjn} \;; \qquad K_j\cdot p_n = - \eta_j \eta_n F_{jji}\;.
\label{kp.all}
\end{align}

Lastly, by substituting \ref{K.lor} in \ref{l2.1}, we have
\begin{align}
&\tilde{\mathcal{L}}_2^{\alpha \beta; G^2} =  
G^2 \omega \sum_{n,i,j} \eta_n H_{nij}\left[(A^i_n + A^n_i) p^{[\alpha}_i p_n^{\beta]} + (A^i_j + A^j_i) p^{[\alpha}_i p_j^{\beta]} + (A^j_n - A^n_j) p^{[\alpha}_j p_n^{\beta]} \right] \,. \label{l2.2}
\end{align}

All coefficients can be determined by using \ref{kp.all} in \ref{Ac.def}, and we find 
\begin{align}
&\tilde{\mathcal{L}}_2^{\alpha \beta; G^2}  =  \frac{G^2 \omega}{2} \sum_{n,i,j} I_{nij} \left( (\eta_i H_{nij} - \eta_n H_{inj}) p_i^{[\alpha} p_n^{\beta]} \left(2F_{nij} m_j (\sigma_{nj}\sigma_{ij} - \sigma_{ni})  - F_{nnj}\frac{m_n m_j}{m_i} (\sigma_{nj}^2 - 1)  \right. \right.  \notag\\
&\left. \left. \qquad \qquad \qquad \qquad - F_{iij} \frac{m_i m_j}{m_n} (\sigma_{ij}^2 - 1) + F_{nni} m_n (\sigma_{ni} \sigma_{nj} - \sigma_{ij}) +  F_{iin} m_i (\sigma_{ni} \sigma_{ij} - \sigma_{nj}) \right) \right. \notag\\
&  \left.  \qquad \qquad \qquad \qquad \qquad + (\eta_j H_{nij} + \eta_n H_{jin})  p_j^{[\alpha} p_n^{\beta]} \left(F_{nni} \frac{m_i m_n}{m_j} (\sigma_{ni}^2 - 1)  - F_{jji} \frac{m_i m_j}{m_n} (\sigma_{ij}^2 - 1)  \right.\right. \notag\\
& \left.  \left.   \qquad \qquad \qquad \phantom{\frac{m_i}{m_n}} \qquad \qquad  - F_{jjn} m_j (\sigma_{nj}\sigma_{ij} - \sigma_{ni}) + F_{nnj} m_n (\sigma_{ni} \sigma_{nj} - \sigma_{ij}) \right) \right)\,,
\label{l2.fin}
\end{align}
where 
\be
I_{nij} = \frac{1}{1 + 2 \sigma_{ni} \sigma_{nj} \sigma_{ij} - \sigma_{ni}^2  - \sigma_{nj}^2 - \sigma_{ij}^2} \,.
\ee

We used the property $p_a^{[\alpha} p_b^{\beta]} = p_{[a}^{\alpha} p_{b]}^{\beta}$ to simplify terms inside the sum of \ref{l2.fin}. We note that the result involves no contribution from external particles involving only Weinberg gravitons, i.e. $p_i^{[\alpha} p_j^{\beta]}$. Our final result for the angular momentum is given by the sum of \ref{l2.fin}, \ref{l1.fin} and \ref{sam.fin}.

\section{Summary and Discussion}

In this paper, we considered the eikonal operator formalism in a soft expansion up to double graviton contributions. This provides a generalization of the leading soft limit from known single-mode gravitational dressings to non-linear graviton modes. Our result followed from the generalized Wilson line formalism for all eikonal amplitudes from dressed Schwinger propagators, which we reviewed in Sec. 2. On substituting two real graviton modes, we find the Weinberg soft graviton factor as the leading term and a subleading double graviton vertex contribution that accounts for the recoil of external particles due to radiative exchanges. We subsequently demonstrated that collinear gravitons provide a gauge invariant dressing, which we consider to be those for real elastic eikonal amplitudes. In Sec. 3, we considered the expectation values of the dressing and their relationship with radiative observables in further detail. The dressing can be expressed as the product of two exponential operators, whose exponents involve a single graviton mode and two graviton modes. The former describes an effective coherent contribution to the dressing, with corrections to all odd powers in the gravitational coupling $\kappa$. All classical observables were shown to result from expectation values with respect to this coherent contribution to the gravitational dressing. In Section 4, we derived the waveform, radiative momentum spectrum and angular momentum, resulting from the  $\kappa^3$ correction in the coherent dressing of the Weinberg soft factor. These observables are linear in $\omega$, a cut-off on the combined frequencies of the two gravitons. While the observables we derive are soft, they are not considered in the strictly soft limit of $\omega \to 0$ and hence do not correspond to `static' observables as with the Weinberg soft factor. Nevertheless, the results we derive for the waveform and angular momentum are sensitive to the $i0$ prescription for the two gravitons in the double graviton factor.

The waveform derived in \ref{mem} results from the expectation value of a single graviton mode operator, and its asymptotic expression for a distant observer was evaluated in a saddle approximation along the direction from the source to the detector. We find a general result in \ref{mem.int7} with a retarded time $u$ dependence. Expanding the waveform in the low-frequency limit and considering $\omega u$ a small constant, we then found a soft waveform in \ref{mem.int8} linear in $\omega$. The asymptotic difference of this waveform from $u \to \infty$ to $u \to -\infty$ provides a $\mathcal{O}(G^2)$ memory effect, which we considered in a collinear limit of the external particles in~\ref{mem.lead}.  We further established that this result agrees with an ultrarelativistic particle approximation of the nonlinear memory effect, in which the differential radiative energy distribution over angles in the usual nonlinear memory effect formula~\ref{gw.nlm} is taken to be that for soft gravitons sourced by ultrarelativistic particles. Apart from realizing the nonlinear memory effect associated with soft gravitons, this demonstrates that the $\mathcal{O}(G^2)$ memory due to the double graviton vertex is associated with the recoil of external particles following the prior emission/absorption of gravitons. It will be interesting to understand the relationship of our $\mathcal{O}(G^2)$ waveform contribution with the recently derived one-loop corrected waveform~\cite{Elkhidir:2023dco,Brandhuber:2023hhy,Herderschee:2023fxh,Georgoudis:2023lgf,Bini:2023fiz} and its analytic expression using the subleading single soft graviton expansion~\cite{Georgoudis:2023eke,Georgoudis:2023ozp,Bini:2024rsy}

We also derived double graviton corrections of the Weinberg soft factor contributions to the radiative momentum spectrum and angular momentum. These observables formally involve a sum over three particles with a double graviton vertex and two single Weinberg soft graviton vertices.  The $\mathcal{O}(G^2)$ radiative momentum spectrum was derived in \ref{emm.g21}. The $\mathcal{O}(G^2)$ correction to the radiative energy spectrum was shown to be related to a corresponding $\mathcal{O}(G^2)$ correction to the imaginary part of the eikonal phase in~\ref{imew}. This provides an extension of its $\mathcal{O}(G)$ version discussed in \cite{DiVecchia:2021ndb, DiVecchia:2022nna}. 
However, as ~\ref{imew} holds outside the strict soft limit, this can also be considered an extension to subleading soft order. We also showed that the radiative angular momentum at $\mathcal{O}(G^2)$ involving the double graviton vertex is given by the gauge invariant combination of the spin angular momentum \ref{sam.fin} and orbital angular momentum contributions \ref{l1.fin} and \ref{l2.fin}, of which \ref{l1.fin} is a mass-dipole orbital angular momentum term. It will be interesting to explore the relationship between the radiative angular momentum at 
$\mathcal{O}(G^2)$ found in this work and the ones in other literature by different approaches \cite{Damour:2020tta, Chen:2021szm, Bini:2022wrq, Manohar:2022dea, Heissenberg:2022tsn, Riva:2023rcm}.

The $\kappa^3$ correction in the coherent dressing results from contracting one of the two gravitons in the double graviton vertex with a single Weinberg soft graviton. Consequently, all $\mathcal{O}(G^2)$ observables considered in Sec. 4 are derived from a coherent operator involving a single graviton. Our results are consistent with the observations in \cite{Cristofoli:2021jas,Britto:2021pud,DiVecchia:2022piu} -- that gravitational waveforms result from coherent states and that the number operator expectation value \ref{poiss} has a leading Poissonian distribution ($ \mathcal{O}(\hbar^{-1})$), with super-Poissonian contributions classically suppressed ($ \mathcal{O}(\hbar^{0})$). As a consequence, the radiated momentum $P^{\alpha}$ is a classical observable since it is related to the number operator by $\hbar k^{\alpha}$. More specific to the eikonal operator formalism, we find that the number operator expectation value \ref{poiss} and radiated energy spectrum \ref{imew} are related to the imaginary part of the eikonal in a similar fashion as~\cite{DiVecchia:2021ndb, DiVecchia:2022nna}. We lastly note that the operator in \ref{dress.fact} can be considered a continuous frequency version of squeezed coherent states, with $\exp[-\Delta_2]$ as the squeezing operator. Notably, squeezed coherent states are minimum uncertainty states with the squeezing operator responsible for super-Poissonian statistics \cite{GerryKinght}, and it would be interesting to consider the implications of such states on the analysis of~\cite{Cristofoli:2021jas}.

\begin{figure}
\begin{center}
\includegraphics[scale=0.7]{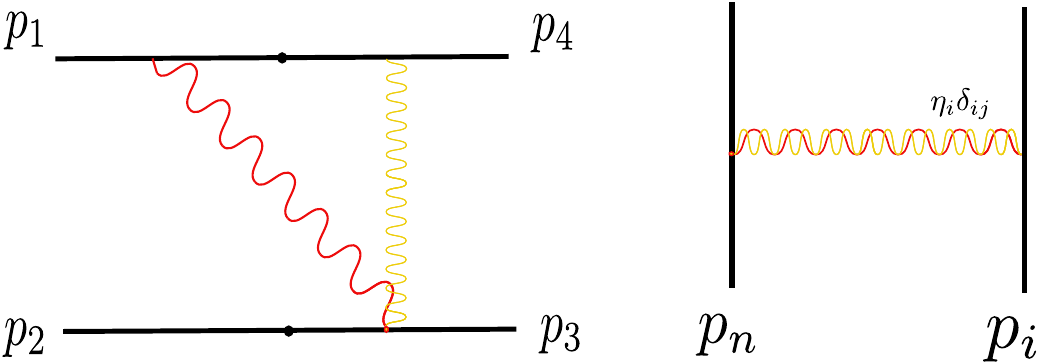}
\end{center}
\caption{Scattering process due to soft graviton exchanges involving a double graviton vertex and two single graviton Weinberg vertices. The particles with a double graviton vertex are labeled by the index $n$ and the ones with a single graviton vertex by $i$ and $j$. The left subfigure is a general diagram with $i\ne j$, which is inconsistent with the collinear requirement on the double graviton vertex. The right subfigure is our resolution by requiring $i=j$, which collapses the two Weinberg vertices into one by implementing an additional factor $\eta_i \delta_{ij}$.   
} \label{fig1}
\end{figure}

Another aspect of our double soft graviton consideration was the collinear condition, which restricted the orientations of gravitons in the double graviton vertex to be the same. By following the conventional prescription of summing soft factors over all external legs, our results, in general, contain an additional sum over all external particles when compared with the corresponding observables in \cite{DiVecchia:2022nna, DiVecchia:2022owy}. Below, we will try to argue that the general sum over external legs for the radiative momentum spectrum and angular momentum in Sec. 4 must be further restricted to be consistent with the collinear emissions from a double graviton vertex. In Fig. \ref{fig1}, we consider an example of the general sum over the indices of the external legs in a 2-2 scattering process. As the two gravitons from the double graviton vertex should be collinear to ensure gauge invariance, this implies that they should also end on the same external leg, which is equivalent to imposing $i=j$. Thus, we shall not treat the indices $n,i$ and $j$ distinct in the sum since they correspond to non-collinear gravitons connected to the double vertex, as illustrated in the left diagram of Fig. \ref{fig1}. Instead, we shall treat $i=j$ so that it leads to a single vertex for two gravitons on both particles, as illustrated in the right diagram of Fig. \ref{fig1}.  Hence, it is more reasonable to require the two soft Weinberg graviton vertices to have a one vertex limit, which can be implemented by including $\eta_i \delta_{ij}$ in the sums over three particle labels. We will explore classical observables in this limit through future work.

There are several avenues for future research. One involves performing the sums described above to derive 4PM and higher gravitational wave observables. The analysis can be carried out along similar lines as those for the Weinberg soft factor dressing for eikonal amplitudes in \cite{DiVecchia:2022piu}. This, in particular, involves substituting expressions for the relative velocities $\sigma_{ab}$ in terms of the PM expansion for the impulse following eikonal kinematics. It will also be interesting to consider the implications of the quadratic graviton mode operator $e^{-\Delta_2}$ in \ref{dress.fact} for gravitationally interacting bodies more generally. While this operator will not contribute to classical radiative observables, it may be interesting to consider in other areas. In this regard, we note that entanglement entropy can be used to constrain scattering amplitudes~\cite{Aoude:2020mlg,Bose:2020shm,Sinha:2022crx,Cheung:2023hkq,Aoude:2024xpx} and more specifically amplitudes with a Weinberg soft graviton dressing~\cite{Carney:2017oxp,Carney:2018ygh,Semenoff:2019dqe}. It will be interesting to investigate dressings with non-linear graviton modes on these results.

\paragraph{Acknowledgement :} 
We would like to thank Paolo Di Vecchia, Carlo Heissenberg, Filippo Vernizzi, Chia-Hsien Shen, Yu-tin Huang and Fei Teng for their valuable feedback and comments on our results. FLL thanks the hospitality of the organizers of the workshop "Gravitational waves meet effective field theories" held at Benasque, Spain on Aug 20-26 of 2023, during which he presented an earlier version of this work. The work of KF is supported by Taiwan's NSTC with grant numbers 111-2811-M-003-005 and 112-2811-M-003 -003-MY3. The work of FLL is supported by Taiwan's NSTC with grant numbers 109-2112-M-003-007-MY3 and 112-2112-M-003-006-MY3. We also like to thank the support of iCAG funding by NTNU.

\bigskip

\bigskip

\appendix

\section{General form of factorized gravitational dressing} \label{A}

This Appendix provides a detailed derivation of \ref{dress.fact} to all orders in $\kappa$. We further consider the evolution of creation and annihilation modes by the dressing operator, which can be used to compute observables.

Given the operator $\exp[-\tilde{\Delta}_1 - \Delta_2]$ we first require that it be equivalent to another operator $\exp[-\bar{\Delta}]\exp[-\Delta_2]$. This implies that
\begin{equation}
\exp[-\bar{\Delta}] =  \exp[-\tilde{\Delta}_1 - \Delta_2] \exp[\Delta_2]\;,
\label{db.exp}
\end{equation}
where $\bar{\Delta}$ can now be determined by using the BCH formula. To simplify notation, we adopt the letter notation for nested commutator, 
\begin{equation}
\left[X_1\,,\left[X_2\,,\left[X_3 \left[X_{4}\,,\left[X_{5}\,, \cdots \right]\right]\right]\right]\right] = \left[X_1 X_2 X_3 X_4 X_5 \cdots \right]\;.
\end{equation}

The BCH formula expanded to the first five orders takes the form~\cite{BCH}
\begin{align}
\ln \left(\exp[X] \exp[Y]\right) & = X + Y + \frac{1}{2}[XY] + \frac{1}{12}[(X-Y)XY] - \frac{1}{24}[XYXY] +\frac{1}{6!} \left( - [X^4Y] - [Y^4X]  \right.\notag\\
&\left. \qquad \quad + 2 [XY^3X] + 2 [YX^3Y] + 6[YXYXY] + 6[XYXYX] \right)+ \cdots\;.
\end{align}
We apply this in \ref{db.exp} by taking $X= -\tilde{\Delta}_1 - \Delta_2$ and $Y = \Delta_2$. This gives
\begin{align}
-\bar{\Delta} &= \ln \left(\exp[-\tilde{\Delta}_1 - \Delta_2] \exp[\Delta_2]\right)\;, \notag\\
& = -\tilde{\Delta}_1 +\frac{1}{2}\left[\Delta_2 \tilde{\Delta}_1\right] - \frac{1}{12} \left[(\tilde{\Delta}_1 + 2 \Delta_2) \Delta_2 \tilde{\Delta}_1\right] + \frac{1}{24} \left[(\tilde{\Delta}_1 +  \Delta_2) \Delta_2^2  \tilde{\Delta}_1\right]\notag\\
& \qquad  +\frac{1}{6!}\left(\left[(\tilde{\Delta}_1 + 2 \Delta_2)^3 \Delta_2 \tilde{\Delta}_1\right] - 6\left[(\tilde{\Delta}_1 + 2 \Delta_2) \Delta_2 (\tilde{\Delta}_1 + 2 \Delta_2) \Delta_2 \tilde{\Delta}_1\right] + 2 \left[(\tilde{\Delta}_1 + 2 \Delta_2) \Delta_2^3 \tilde{\Delta}_1\right]  \right. \notag\\
&\left. \qquad \qquad \quad  + 2 \left[\Delta_2 (\tilde{\Delta}_1 + 2 \Delta_2)^2 \Delta_2 \tilde{\Delta}_1\right] - 6\left[\Delta_2 (\tilde{\Delta}_1 + 2 \Delta_2) \Delta_2^2 \tilde{\Delta}_1\right] + \left[\Delta_2^4 \tilde{\Delta}_1\right] \right) + \cdots\;.
\label{bch.del}
\end{align}
The series in \ref{bch.del} simplifies into a sum over two terms. Those with $\left[\Delta_2^n \tilde{\Delta}_1\right]$ for all $n\ge 0$, which is a single graviton state of the form $\tilde{\Delta}_1$ but with different coefficients for the graviton creation and annihilation modes. The only contributions involving more than one $\tilde{\Delta}_1$ in the nested commutators is of the form  $\left[\tilde{\Delta}_1 \Delta_2^n \tilde{\Delta}_1\right]$, which involve no graviton modes. We find that \ref{bch.del} can be expressed as
\begin{align}
-\bar{\Delta} &= - \Delta_1 - \Delta_0 \notag\\
\Delta_1& = \tilde{\Delta}_1 + \sum_{n=1}^{\infty} \frac{(-1)^n}{(n+1)!} \left[\Delta_2^n \tilde{\Delta}_1\right] \label{sg.fact}\\
\Delta_0 & = \sum_{n=1}^{\infty} c_n \frac{(-1)^{n+1}}{(n+2)!} \left[\tilde{\Delta}_1 \Delta_2^n \tilde{\Delta}_1\right] \,,\label{zg.fact}
\end{align}
where we have explicitly separated the $n=0$ contribution in $\Delta_1$, which is the single graviton term of the unfactorized gravitational dressing.  We hence find that $-\bar{\Delta}$ can be written as a sum over a single graviton mode and a zero graviton factor known up to certain constants $c_n$, with the first few constants being $c_1 = \frac{1}{2}\,, c_2 = 1 \,, c_{3} = \frac{1}{2}\,, c_{4} = \frac{3}{2}\,, \cdots$. However, in the following, we show that $\Delta_0$ vanishes, and hence the dressing $\exp[-\bar{\Delta}]$ can be written as a normalized single graviton dressing. We also note that since $\tilde{\Delta}_1$ and $\Delta_2$ are respectively ${\cal O}(\kappa)$ and ${\cal O}(\kappa^2)$, \ref{sg.fact} provides a single graviton (coherent) dressing to ${\cal O}(\kappa^{2n+1})$ for all positive integers $n$.

Using the expressions for $\Delta_2$ and $\tilde{\Delta}_1$, we determine the following result for $\Delta_1$ and $\Delta_0$
\begin{align}
\Delta_1 &= \frac{1}{\hbar} \int_{\vec{k}} d^3k \left(a_i (k) f^*_i(k) - a^{\dagger}_i (k) f_i(k) \right) \notag\\
&\qquad +  \frac{1}{\hbar} \int_{\vec{l}} d^3l \int_{\vec{k}} d^3k a_i (k)\left[ \sum_{n=1}^{\infty} \frac{1}{(n+1)!} \left(\mathcal{C}_{ij;n}^{*} (k\,,l) f^*_{j}(l) -  \mathcal{S}_{ij;n}^{*} (k\,,l) f_{j}(l) \right)\right] \notag\\
& \qquad \qquad -  \frac{1}{\hbar} \int_{\vec{l}} d^3l \int_{\vec{k}} d^3k a^{\dagger}_i (k)\left[ \sum_{n=1}^{\infty} \frac{1}{(n+1)!} \left(\mathcal{C}_{ij;n} (k\,,l) f_{j}(l) - \mathcal{S}_{ij;n}(k\,,l)  f^*_{j}(l)\right)\right] \,, \label{d1.f2}\\
\Delta_0 & = \frac{1}{\hbar} \int_{\vec{l}} d^3l \int_{\vec{k}} d^3k \left[ \sum_{n=1}^{\infty} \frac{c_n}{(n+2)!} \left(f^*_i(k)\mathcal{C}_{ij;n}(k\,,l) f_{j}(l) - f_i(k)\mathcal{C}_{ij;n}^{*} (k\,,l) f^*_{j}(l) \phantom{\sum_{n} \frac{c_n}{(n+2)!}} \right.\right. \notag\\
& \left.\left. \quad \qquad \qquad \qquad \qquad \phantom{\sum_{n} \frac{c_n}{(n+2)!}} + f_i(k) \mathcal{S}_{ij;n}^{*} (k\,,l) f_{j}(l) - f^*_i(k) \mathcal{S}_{ij;n}(k\,,l) f^*_{j}(l) \right)\right]  \label{d0}
\end{align}
with $\mathcal{S}_{ij;n}(k\,,l)$ and $\mathcal{C}_{ij;n}(k\,,l)$ defined in terms of a products over an integrated expression $I^{(\pm) \mu_m \nu_m}_{\mu_{m+1} \nu_{m+1}} (l_{m+1})$
\begin{align}
S_{ij;n}(k\,,l) &= \frac{1}{2}\int_{\vec{l}_n} d^3l_n\, \varepsilon^*_{i\,,\mu_0 \nu_0} (l_0) \, \varepsilon_j^{*\;\mu_n \nu_n} (l_n) \, \delta(\vec{l_0}\,, \vec{k})\, \delta(\vec{l_n}\,, \vec{l}) \notag\\
& \qquad \qquad   \left[ \left(\prod_{m=0}^{n-1} I^{(-) \mu_m \nu_m}_{\mu_{m+1} \nu_{m+1}} (l_{m+1})\right) +  (-1)^{n+1}\left(\prod_{m=0}^{n-1} I^{(+) \mu_m \nu_m}_{\mu_{m+1} \nu_{m+1}} (l_{m+1})\right)\right]\;,  \label{s.def}\\
C_{ij;n}(k\,,l)  &= \frac{1}{2}\int_{\vec{l}_n} d^3l_n \, \varepsilon^*_{i\,,\mu_0 \nu_0} (l_0)\, \varepsilon_j^{\mu_n \nu_n} (l_n) \,\delta(\vec{l_0}\,, \vec{k})\,\delta(\vec{l_n}\,, \vec{l}) \notag\\
& \qquad \qquad \left[ \left(\prod_{m=0}^{n-1} I^{(-) \mu_m \nu_m}_{\mu_{m+1} \nu_{m+1}} (l_{m+1})\right) +  (-1)^{n}\left(\prod_{m=0}^{n-1} I^{(+) \mu_m \nu_m}_{\mu_{m+1} \nu_{m+1}} (l_{m+1})\right)\right]\;,  \label{c.def}\\
I^{(\pm) \mu_m \nu_m}_{\mu_{m+1} \nu_{m+1}} (l_{m+1}) &= \int_{\vec{l}_m} d^3l_m\,  \Pi^{\mu_m \nu_m}_{\alpha_{m} \beta_{m}}(l_m) \Pi^{\alpha_{m+1} \beta_{m+1}}_{\mu_{m+1} \nu_{m+1}}(l_{m+1}) \notag\\
& \qquad \qquad \qquad \left[A^{\alpha_m \beta_m}_{\alpha_{m+1} \beta_{m+1}}(l_m\,,l_{m+1}) \pm B^{\alpha_m \beta_m}_{\alpha_{m+1} \beta_{m+1}}(l_m\,,l_{m+1})\right]\;. \label{I.def}
\end{align}

As the integrals in \ref{d0} are not sensitive to the poles of $\omega_k$ and $\omega_l$, we have $\Delta_0 = 0$. We conclude with the general expressions for $e^{\Delta} a_i(k) e^{-\Delta}$ and $e^{\Delta} a_{i}^{\dagger} (k) e^{-\Delta}$ to all orders in $\kappa$
\begin{align}
e^{\Delta}a_i(k) e^{-\Delta} &=  a_i(k) + f_i(k) \notag\\
& \qquad  + \frac{1}{\hbar} \int_{\vec{l}} d^3l \int_{\vec{k}} d^3k \left[ \sum_{n=1}^{\infty} \frac{1}{(n+1)!} \left(\mathcal{C}_{ij;n} (k\,,l) f_{j}(l) - \mathcal{S}_{ij;n}(k\,,l)  f^*_{j}(l)\right)\right]\notag\\
& \qquad \qquad \quad +  \frac{1}{\hbar} \int_{\vec{l}} d^3l \int_{\vec{k}} d^3k \left[ \sum_{n=1}^{\infty} \frac{1}{n!} \left(\mathcal{C}_{ij;n} (k\,,l) a_{j}(l) - \mathcal{S}_{ij;n}(k\,,l)  a^{\dagger}_{j}(l)\right)\right] \label{ag2.evg}\\
 e^{\Delta}a^{\dagger}_i(k) e^{-\Delta} &= a^{\dagger}_i(k) + f^*_i(k) \notag\\
& \qquad + \frac{1}{\hbar} \int_{\vec{l}} d^3l \int_{\vec{k}} d^3k \left[ \sum_{n=1}^{\infty} \frac{1}{(n+1)!} \left(\mathcal{C}_{ij;n}^{*} (k\,,l) f^*_{j}(l) -  \mathcal{S}_{ij;n}^{*} (k\,,l) f_{j}(l) \right)\right] \notag\\
& \qquad  \qquad \quad +   \frac{1}{\hbar} \int_{\vec{l}} d^3l \int_{\vec{k}} d^3k \left[ \sum_{n=1}^{\infty} \frac{1}{n!} \left(\mathcal{C}_{ij;n}^{*} (k\,,l) a^{\dagger}_{j}(l) -  \mathcal{S}_{ij;n}^{*} (k\,,l) a_{j}(l) \right)\right]\;. \label{adg2.evg}
\end{align}
This is the all order generalization of the $n=1$ (${\cal O}(\kappa^3)$) correction result in \ref{ag2.ev1} and \ref{adg2.ev1}.

\section{Evaluation of $F_{inm}$ and $\vec{G}_{inm}$} \label{B}
In the following, we will always consider the graviton orientation to be given by
\begin{equation}
\hat{k} = \left( \sqrt{1 - y^2} \cos \phi \,, \sqrt{1 - y^2} \sin \phi \,, y\right)\,,
\label{k.p}
\end{equation}
with $y = \cos \theta$ . The parameterizations adopted for $\vec{p}_n \,, \vec{p}_m$ and $\vec{p}_i$ in the two subsections will be treated differently. We will however always use the freedom to have all the hard particle momenta in a $2$-dimensional plane, which we consider along the `x-z' plane throughout. The integrals we consider will also have a non-trivial integration over $\phi$, and will make use of the following result
\begin{equation}
\int_{0}^{2\pi}d\phi \frac{1}{A + B \cos \phi} = \frac{2\pi}{\sqrt{A^2 - B^2}}\;. \label{phi.int}
\end{equation}

\subsection{Evaluation of $F_{inm}$}
We consider the following parametrization for the external particle spatial momenta
\begin{align}
\vec{p}_i = \left(0\,,0\,, \bar{p}_i\right) \;,\;&\;  \vec{p}_n = \left(-\bar{p}_j\,,0\,,- \bar{p}_i\right)\;,  \;\; \vec{p}_j = \left(\bar{p}_j \,,0\,, 0\right)   \;,\label{sp.p}
\end{align}
which satisfy the conservation of momentum: $\vec{p}_n + \vec{p}_i + \vec{p}_j = 0$

We then have the following integral for \ref{ang.f}
\begin{align}
F_{nij} & := \frac{1}{2 \pi}\oint d\Omega_k  \frac{1}{(E_n - \vec{p}_n.\hat{k})(E_i - \vec{p}_i.\hat{k})(E_j - \vec{p}_j.\hat{k})}  \,,\notag\\
&= \frac{1}{2 \pi}\int_{-1}^{1}dy  \frac{1}{(E_i - \bar{p}_i y)} \int_{0}^{2\pi}d\phi \frac{1}{(E_n + \bar{p}_i y + \bar{p}_j\sqrt{1-y^2} \cos \phi)(E_j - \bar{p}_j\sqrt{1-y^2} \cos \phi)}\notag\\
&= \frac{1}{2 \pi}\int_{-1}^{1}dy  \frac{1}{(E_i - \bar{p}_i y) (E_n + E_j + \bar{p}_i y )}\notag\\
&\qquad \qquad \int_{0}^{2\pi}d\phi \left[\frac{1}{E_n + \bar{p}_i y + \bar{p}_j\sqrt{1-y^2} \cos \phi} + \frac{1}{E_j - \bar{p}_j\sqrt{1-y^2} \cos \phi} \right]
\label{fint.1}
\end{align}

Applying \ref{phi.int} in \ref{fint.1} gives
\begin{align}
F_{nij} & = \int_{-1}^{1}dy  \frac{1}{(E_i - \bar{p}_i y) (E_n + E_j + \bar{p}_i y )} \left[\frac{1}{\sqrt{(E_n + \bar{p}_i y)^2 - \bar{p}_j^2 (1-y^2)}} + \frac{1}{\sqrt{E_j^2 - \bar{p}_j^2 (1-y^2)}} \right]\notag\\
& = \frac{1}{E_n + E_i + E_j} \left[\frac{1}{\sqrt{ E_i^2 E_j^2 - m_i^2 m_j^2}}\ln \left(\frac{E_i E_j +\sqrt{ E_i^2 E_j^2 - m_i^2 m_j^2}}{E_i E_j  - \sqrt{ E_i^2 E_j^2 - m_i^2 m_j^2}} \right) \right. \notag\\
& \left. \qquad + \frac{1}{\sqrt{ (E_n +  E_i)^2 \bar{p}_i^2 + m_i^2 \bar{p}_j^2}} \ln \left(\frac{E_n E_i + \bar{p}_i^2 + \sqrt{ (E_n +  E_i)^2 \bar{p}_i^2 + m_i^2 \bar{p}_j^2}}{E_n E_i + \bar{p}_i^2 - \sqrt{ (E_n +  E_i)^2 \bar{p}_i^2 + m_i^2 \bar{p}_j^2}} \right) \right. \notag\\
& \left. \qquad \quad + \frac{1}{\sqrt{ (E_n +  E_j)^2 \bar{p}_j^2 + m_j^2 \bar{p}_i^2}} \ln \left(\frac{E_n E_j + \bar{p}_j^2 + \sqrt{ (E_n +  E_j)^2 \bar{p}_j^2 + m_j^2 \bar{p}_i^2}}{E_n E_j + \bar{p}_j^2 - \sqrt{ (E_n +  E_j)^2 \bar{p}_j^2 + m_j^2 \bar{p}_i^2}} \right) \right]
\label{fint.2}
\end{align}

From \ref{rv.def} and $-p_a^2 = E_a^2 - \vec{p}_a^2 = m_a^2$ for $a= n,i,j$, we then find that the parametrization \ref{sp.p} gives the following relations
\begin{align}
E_i E_j = m_i m_j \sigma_{ij} &\,, \qquad \qquad E_i^2 E_j^2 - m_i^2 m_j^2 = m_i^2 m_j^2 \left(\sigma_{ij}^2 - 1\right)\;, \notag\\
 E_n E_j + \bar{p}_j^2 = m_n m_j \sigma_{nj} &\,, \qquad \qquad  (E_n +  E_j)^2 \bar{p}_j^2 + m_j^2 \bar{p}_i^2 = m_n^2 m_j^2 \left(\sigma_{nj}^2 - 1\right)\;,  \notag\\
E_n E_i + \bar{p}_i^2 = m_n m_i \sigma_{ni} &\,, \qquad \qquad (E_n +  E_i)^2 \bar{p}_i^2 + m_i^2 \bar{p}_j^2 = m_n^2 m_i^2 \left(\sigma_{ni}^2 - 1\right)\;,  \notag\\E_n + E_i + E_j & = \sqrt{2 (m_i m_j \sigma_{ij}+ m_n m_j \sigma_{nj} + m_n m_i \sigma_{ni}) + m_i^2  + m_j^2 + m_n^2}\;.
\label{inv.1}
\end{align}

Substituting \ref{inv.1} in \ref{fint.2}, we get the result 
\begin{align}
F_{nij} & = \frac{2\; (m_n\Delta_{ij}+ m_i \Delta_{nj}+m_j \Delta_{ni}) }{m_n m_i m_j\sqrt{2 (m_i m_j \sigma_{ij}+ m_n m_j \sigma_{nj} + m_n m_i \sigma_{ni}) + m_i^2  + m_j^2 + m_n^2}} \;,
\end{align}
where $\Delta_{ab}$ is defined in \ref{rv.def}. 

\subsection{Evaluation of $\vec{G}_{inm}$} \label{Appendix.G}

We now consider the derivation of \ref{ang.g}. Start with the following decomposition:
\begin{equation}
\vec{G}_{nij} := {1\over 2\pi} \oint d\Omega_k  \frac{\hat{k}}{(E_i - \vec{p}_i.\hat{k})(E_n - \vec{p}_n.\hat{k})(E_j - \vec{p}_m.\hat{k})} := a_{nij} \hat{p}_i + b_{nij} \hat{p}_j +  c_{nij} \hat{p}_n\,.
\label{gi.ri}
\end{equation}
In the above, we used the rotational symmetry of the integral to express it as a linear combination over the hard particle orientations in the last equality. Defining $\hat{p}_a \cdot \hat{p}_b = \cos \theta_{ab}$ as the relative angle between two hard particle orientations $\hat{p}_a$ and $\hat{p}_b$, we then find the following solutions for $a_{nij}\,, b_{nij}$ and $c_{nij}$ from \ref{gi.ri}
\begin{align}
a_{nij} & = \frac{\vec{G} \cdot \hat{p}_i \sin^2 \theta_{nj} + \cos \theta_{nj}\left(\vec{G} \cdot\hat{p}_j \cos \theta_{ni} + \vec{G}\cdot\hat{p}_n \cos \theta_{ij}\right) - \left(\vec{G}\cdot\hat{p}_n \cos \theta_{in} + \vec{G}\cdot\hat{p}_j \cos \theta_{ij}\right)}{1 + 2 \cos \theta_{ij} \cos \theta_{ni}\cos \theta_{nj} - \cos^2 \theta_{ij} - \cos^2 \theta_{ni}  - \cos^2 \theta_{nj}} \notag\\
b_{nij} & = \frac{\vec{G}\cdot\hat{p}_j \sin^2 \theta_{ni} + \cos \theta_{ni}\left(\vec{G}\cdot \hat{p}_i \cos \theta_{nj} + \vec{G} \cdot \hat{p}_n \cos \theta_{ij}\right) - \left(\vec{G}\cdot \hat{p}_n \cos \theta_{nj} + \vec{G}\cdot \hat{p}_i \cos \theta_{ij}\right)}{1 + 2 \cos \theta_{ij} \cos \theta_{ni}\cos \theta_{nj} - \cos^2 \theta_{ij} - \cos^2 \theta_{ni}  - \cos^2 \theta_{nj}} \notag\\
c_{nij} & = \frac{\vec{G}\cdot \hat{p}_n \sin^2 \theta_{ij} + \cos \theta_{ij}\left(\vec{G} \cdot \hat{p}_i \cos \theta_{nj} + \vec{G}\cdot \hat{p}_j \cos \theta_{ni}\right) - \left(\vec{G} \cdot \hat{p}_j \cos \theta_{nj} + \vec{G} \cdot \hat{p}_i \cos \theta_{ni}\right)}{1 + 2 \cos \theta_{ij} \cos \theta_{ni}\cos \theta_{nj} - \cos^2 \theta_{ij} - \cos^2 \theta_{ni}  - \cos^2 \theta_{nj}} \label{coeff.def}
\end{align} 
In the above, we have omitted the sub-indices of $\vec{G}_{nij}$ to avoid clutter of expressions.

Thus, the expression for $\vec{G}_{nij}$ is completely specified from its contraction with the hard particle orientations. In the following, we consider the derivation for 

\begin{align}
\vec{G}_{nij}\cdot \hat{p}_a = \frac{1}{2 \pi} \oint d\Omega_k  \frac{\hat{p}_a \cdot \hat{k}}{(E_n - \vec{p}_n\cdot \hat{k})(E_i - \vec{p}_i\cdot \hat{k})(E_j - \vec{p}_j \cdot \hat{k})} \,, \label{gpa.int}
\end{align}

which cover the cases for $a = n,i,j$ by an appropriate choice of indices. We can carry out the integration using the parametrization 
\begin{align}
\vec{p}_i = \bar{p}_i \hat{p}_i \;; & \qquad  \hat{p}_i = \left(0\,,0\,, 1\right) \,,\notag\\
\vec{p}_n = \bar{p}_n \hat{p}_n \;; & \qquad  \hat{p}_n = \left(\sin \theta_{ni} \,,0\,, \cos \theta_{ni}\right) \,, \notag\\
\vec{p}_j = \bar{p}_j \hat{p}_j \;; & \qquad  \hat{p}_j = \left(\sin \theta_{ij} \,,0\,, \cos \theta_{ij}\right) \,,\label{sp.p2}
\end{align}

With this parametrization, we have the following Lorentz invariant quantities
\begin{align}
&\sigma_{ab} = \frac{E_a E_b - \bar{p}_a\bar{p}_b \cos \theta_{ab}}{m_a m_b} \,,  \quad \sigma_{ac} = \frac{E_a E_c - \bar{p}_a\bar{p}_c \cos \theta_{ac}}{m_a m_c} \,, \qquad \sigma_{bc} = \frac{E_b E_c - \bar{p}_b\bar{p}_c \cos (\theta_{ac} - \theta_{ab})}{m_b m_c} \,,\notag\\
&m_a^2 m_c^2 \left(\sigma_{ac}^2 - 1\right) = E_a^2\bar{p}_c^2 +  E_c^2\bar{p}_a^2 -2 E_a E_c \bar{p}_a \bar{p}_c \cos \theta_{ac} - \bar{p}_a^2 \bar{p}_c^2 \sin^2 \theta_{ac} \,, \notag\\
&m_a^2 m_b^2 \left(\sigma_{ab}^2 - 1\right) = E_a^2\bar{p}_b^2 +  E_b^2\bar{p}_a^2 -2 E_a E_b \bar{p}_a \bar{p}_b \cos \theta_{ab} - \bar{p}_a^2 \bar{p}_b^2 \sin^2 \theta_{ab} \,, \notag\\
&m_b^2 m_c^2 \left(\sigma_{bc}^2 - 1\right) = E_c^2\bar{p}_b^2 +  E_b^2\bar{p}_c^2 -2 E_b E_c \bar{p}_b \bar{p}_c \cos (\theta_{ac} - \theta_{ab}) - \bar{p}_b^2 \bar{p}_c^2 \sin^2 (\theta_{ac} - \theta_{ab}) \,. \label{inv.2}
\end{align}

Using \ref{sp.p2} and \ref{k.p} in \ref{gpa.int}, we have

\begin{align}
\vec{G}_{nij}\cdot\hat{p}_a & = \frac{1}{\pi} \int_{-1}^{1}dy  \frac{y}{(E_a - \bar{p}_a y)} \notag\\
&\quad \int_{0}^{\pi}d\phi \frac{1}{(E_b - \bar{p}_b ( \cos \theta_{ab} y + \sin \theta_{ab} \sqrt{1-y^2} \cos \phi))(E_c - \bar{p}_c ( \cos \theta_{ac} y + \sin \theta_{ac} \sqrt{1-y^2} \cos \phi))}\notag\\
&= \int_{-1}^{1}dy  \frac{y}{(E_a - \bar{p}_a y) ((E_b \bar{p}_c \sin \theta_{ac} - E_c \bar{p}_b \sin \theta_{ab}) + y \bar{p}_c \bar{p}_b (\cos \theta_{ac} \sin \theta_{ab} - \cos \theta_{ab} \sin \theta_{ac}) )}\notag\\
&\left[\frac{\bar{p}_c \sin \theta_{ac}}{(E_c - \bar{p}_c \cos \theta_{ac} y)^2 - \bar{p}^2_c (1-y^2) \sin^2 \theta_{ac}} - \frac{\bar{p}_b \sin \theta_{ab}}{(E_b - \bar{p}_b \cos \theta_{ab} y)^2 - \bar{p}^2_b (1-y^2) \sin^2 \theta_{ab}} \right]
\label{gpi.1}
\end{align}

where we used \ref{phi.int} to arrive at the last line of \ref{gpi.1}. The $y$ integral can be evaluated to find a result that depends on the Lorentz invariant quatities, as well as on $\{\frac{E_a}{m_a}\,,\frac{E_b}{m_b}\,,\frac{E_c}{m_c}\}$ and $\frac{\bar{p}_a}{m_a}$. Defining 
\begin{align}
\gamma^{(a)}_{bc} := \frac{m_a}{\bar{p}_a}\left(\frac{E_b}{m_{b}}\sigma_{ab} - \frac{E_c}{m_{c}}\sigma_{ac}\right)\,, &\qquad \chi^{(a)}_{a b} := \sqrt{\sigma_{ab}^2 - 1 + \left(\frac{E_b^2}{m_b^2}+ \frac{E_a^2}{m_a^2} - 2 \frac{E_a}{m_a}\frac{E_b}{m_b}\sigma_{ab}\right)} \,, \notag\\
\Delta^{(a)}_{a b c} &:= \chi^{(a)}_{a c} \gamma^{(a)}_{ba} - \chi^{(a)}_{a b} \gamma^{(a)}_{ca} \,,\notag\\
\alpha^{(a)}_{bc}:= \sigma_{bc} - \frac{\chi^{(a)}_{a b}}{\chi^{(a)}_{a c}} \,,& \qquad \beta^{(a)}_{bc}:= \sigma_{bc} - \frac{\chi^{(a)}_{a c}}{\chi^{(a)}_{a b}} \,; \qquad \qquad (b\,,c \neq a) \,,\notag\\
\xi^{(a)}_{ab} := \frac{\sigma_{ab}^2 - 1 + \gamma^{(a)}_{b a}\sqrt{\sigma_{ab}^2 - 1}}{\sigma_{ab}\left(\gamma^{(a)}_{b a} + \sqrt{\sigma_{ab}^2 - 1}\right)}\,, &\qquad \xi^{(a)}_{ac} := \frac{\sigma_{ac}^2 - 1 + \gamma^{(a)}_{c a}\sqrt{\sigma_{ac}^2 - 1}}{\sigma_{ac}\left(\gamma^{(a)}_{c a} + \sqrt{\sigma_{ac}^2 - 1}\right)}\,, \notag\\
\mu^{(a)}_{bc} := \frac{\sigma_{bc}^2 - 1 + \gamma^{(a)}_{bc}\sqrt{\sigma_{bc}^2 - 1}}{\alpha^{(a)}_{bc}\left(\gamma^{(a)}_{bc} + \sqrt{\sigma_{bc}^2 - 1}\right)} \,,& \qquad \nu^{(a)}_{bc} := \frac{\sigma_{bc}^2 - 1 + \gamma^{(a)}_{bc}\sqrt{\sigma_{bc}^2 - 1}}{\beta^{(a)}_{bc}\left(\gamma^{(a)}_{bc} + \sqrt{\sigma_{bc}^2 - 1}\right)}\,,
\label{ept.def}
\end{align}

we find that \ref{gpi.1} has the result

\begin{align}
\vec{G}_{nij}\cdot \hat{p}_a &= \frac{2}{m_a m_b m_c}\frac{1}{ \Delta^{(a)}_{abc}} \left[\chi^{(a)}_{ac} \left(\frac{E_a}{m_a} \frac{\arctanh \xi^{(a)}_{ac}}{\sqrt{\sigma_{ac}^2 - 1}} - \frac{E_b}{m_b}  \frac{\arctanh \mu^{(a)}_{bc} + \arctanh \nu^{(a)}_{bc}}{\sqrt{\sigma_{bc}^2 - 1}}\right)  \right. \notag\\ 
& \left. \quad \qquad \qquad  \qquad  \qquad  -\chi^{(a)}_{ab} \left(\frac{E_a}{m_a} \frac{\arctanh \xi^{(a)}_{ab}}{\sqrt{\sigma_{ab}^2 - 1}} - \frac{E_c}{m_c}  \frac{\arctanh \mu^{(a)}_{bc} + \arctanh \nu^{(a)}_{bc}}{\sqrt{\sigma_{bc}^2 - 1}}\right) \right]\;.
\end{align}

\bibliographystyle{jhep}

\bibliography{refs.bib}	

\end{document}